\documentclass[12pt]{article}
\usepackage{epsf}
\usepackage{epsfig}
\def\beq{\begin{eqnarray}}
\def\eeq{\end{eqnarray}}

\def\no{\nonumber}

\newcommand\BBbar{$B_d^0$--$\overline{B_d^0}$\ }
\newcommand\cO{{\cal O}}

\newcommand{\lsim}{\stackrel{<}{_\sim}}
\newcommand{\gsim}{\stackrel{>}{_\sim}}

\begin{document}
%%%%%%%%%%%%%%%%%%%%%%%%%%%%%%%%%%%%%%%%%%%%%%%%%%%%%%%%%%%%%%%%%%%%%%%%%%%%

\begin{titlepage}

\begin{flushright}
CERN-TH/2003-039\\
UAB-FT-541\\
hep-ph/0302229
\end{flushright}

\vspace{1cm}
\begin{center}
\boldmath
\large\bf

Shedding Light on the ``Dark Side'' of $B^0_d$--$\overline{B^0_d}$ Mixing

\vspace{0.2 truecm}

through $B_d\to\pi^+\pi^-$, $K\to\pi\nu\overline{\nu}$ and 
$B_{d,s}\to\mu^+\mu^-$

\unboldmath
\end{center}

\vspace{0.5cm}

\begin{center}
Robert Fleischer,${}^a$  Gino Isidori,${}^b$  Joaquim Matias${}^c$

\vspace{0.9truecm}

${}^a$ {\sl Theory Division, CERN, CH-1211 Geneva 23, Switzerland}

\vspace{0.2truecm}

${}^b$ {\sl INFN, Laboratori Nazionali di Frascati, I-00044 Frascati, Italy}

\vspace{0.2truecm}

${}^c$ {\sl IFAE, Universitat Aut\`onoma de Barcelona, 08193 Bellaterra,
Barcelona, Spain}
\end{center}

\vspace{0.8cm}
\begin{abstract}
\vspace{0.2cm}\noindent
In a wide class of new-physics models, which can be motivated through generic
arguments and within supersymmetry, we obtain large contributions to
$B^0_d$--$\overline{B^0_d}$ mixing, but not to $\Delta B=1$ processes.
If we assume such a scenario, the solutions $\phi_d\sim47^\circ\lor
133^\circ$ for the $B^0_d$--$\overline{B^0_d}$ mixing phase implied by
${\cal A}_{\rm CP}^{\rm mix}(B_d\to J/\psi K_{\rm S})$ cannot be converted
directly into a constraint in the $\overline{\rho}$--$\overline{\eta}$ plane.
However, we may complement $\phi_d$ with $|V_{ub}/V_{cb}|$ and
the recently measured CP asymmetries in $B_d\to\pi^+\pi^-$ to
determine the unitarity triangle, with its angles $\alpha$, $\beta$ and
$\gamma$. To this end, we have also to control penguin effects, which we do
by means of the CP-averaged $B_d\to\pi^\mp K^\pm$ branching ratio.
Interestingly, the present data show a perfectly consistent picture not only
for the ``standard'' solution of $\phi_d\sim 47^\circ$, but also for
$\phi_d\sim 133^\circ$. In the latter case, the preferred region for the apex
of the unitarity triangle is in the second quadrant, allowing us to accommodate
conveniently $\gamma>90^\circ$, which is also favoured by other non-leptonic
$B$ decays such as $B\to\pi K$. Moreover, also the prediction for
BR$(K^+\to\pi^+\nu\overline{\nu})$ can be brought to better
agreement with experiment. Further strategies to explore this scenario with
the help of $B_{d,s}\to\mu^+\mu^-$ decays are discussed as well.
\end{abstract}

\vfill
\noindent

February 2003

\end{titlepage}

\thispagestyle{empty}
\vbox{}
\newpage

\setcounter{page}{1}

\section{Introduction}\label{intro}
\begin{figure}[t]
\vspace*{-0.5truecm}
\begin{minipage}[t]{75mm}
      \begin{center}
      \epsfxsize=7.0 truecm
      \epsffile{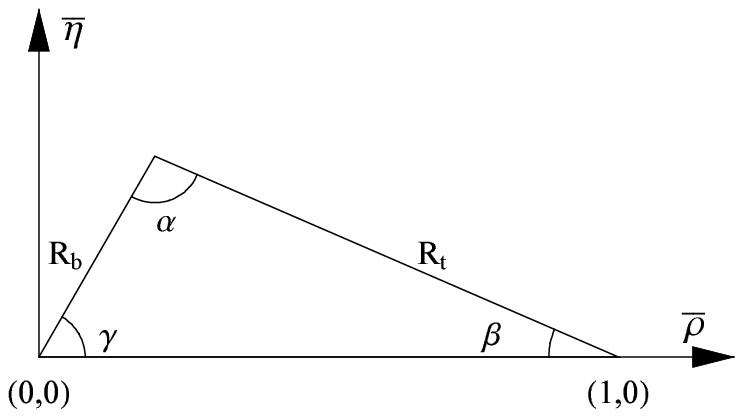}
      {\bf (a) }
      \end{center}
\end{minipage}
\hspace{\fill}
\begin{minipage}[t]{75mm}
      \begin{center}
      \epsfxsize=7.0 truecm
      \epsffile{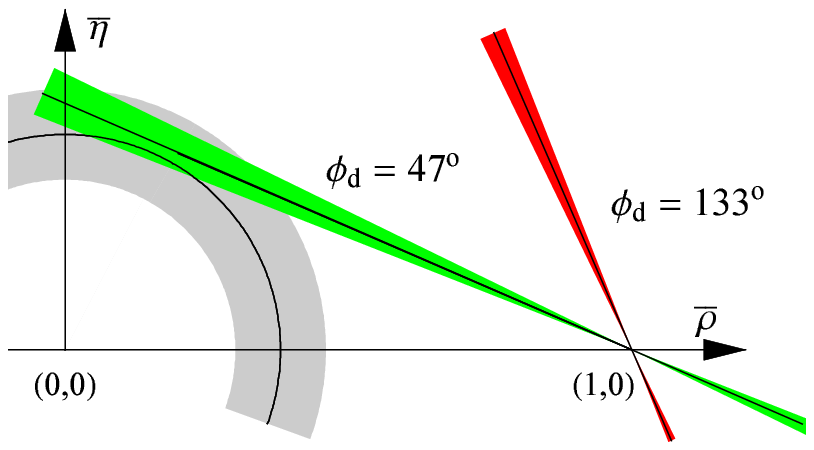}
      {\bf (b) }
      \end{center}
\end{minipage}
\caption{ (a) Definition of the CKM unitarity triangle. (b)
Standard procedure adopted in the literature to represent
the two solutions for $\phi_d$ in (\ref{phid-exp}); as emphasized in
Section~\ref{sec:phid}, this procedure is actually very misleading
in the case of the second ``unconventional'' $\phi_d=133^\circ$ branch.}
\label{fig:UT}
\end{figure}

Thanks to the efforts at the $B$ factories, the exploration of CP violation
is now entering another exciting stage, allowing us to confront the
Kobayashi--Maskawa mechanism \cite{KM} with data. After the discovery
of mixing-induced CP violation in the ``gold-plated'' mode
$B_d\to J/\psi K_{\rm S}$ \cite{CP-B-obs},
as well as important other measurements, one of
the most interesting questions is now to what extent the possible space for
new physics (NP) has already been reduced. In this context, the central target
is the unitarity triangle of the Cabibbo--Kobayashi--Maskawa (CKM) matrix
illustrated in Fig.~\ref{fig:UT}(a), where $\overline{\rho}$ and
$\overline{\eta}$ are the generalized  Wolfenstein parameters
\cite{BLO}. The usual fits for the allowed region for the
apex of the unitarity triangle in the $\overline{\rho}$--$\overline{\eta}$
plane  -- the ``CKM fits'' -- seem to indicate that no NP is
required to accommodate the data \cite{CKM-fits,nir}. However, this is not
the complete answer to this exciting question. In order to fully address it,
the interpretation of the data on the mixing-induced
CP asymmetry\footnote{For a detailed discussion, see \cite{RF-PHYS-REP}.}
\begin{equation}\label{Amix-BdpsiKS}
{\cal A}_{\rm CP}^{\rm mix}(B_d\to J/\psi K_{\rm S})=-\sin\phi_d,
\end{equation}
where $\phi_d$ is the CP-violating weak $B^0_d$--$\overline{B^0_d}$
mixing phase, requires a more involved analysis. Within the Standard Model
(SM), $\phi_d$ is given by $2\beta$. However, due to the possible impact of
physics beyond the SM, $\phi_d$ takes the following general form:
\begin{equation}\label{phid-NP}
\phi_d=\phi_d^{\rm SM} + \phi_d^{\rm NP}= 2\beta + \phi_d^{\rm NP},
\end{equation}
where $\phi_d^{\rm NP}$ is due to non-standard contributions to
$B^0_d$--$\overline{B^0_d}$ mixing, which is the preferred mechanism
for NP to manifest itself in (\ref{Amix-BdpsiKS}).
In principle, physics beyond the SM may also affect the $B\to J/\psi K$ decay
amplitudes; however, in this case the new contributions have to compete
with SM tree-level amplitudes and their relative impact is expected to be
much smaller. So far, a set of ``smoking-gun'' observables to search for
such kind of NP does not indicate any deviation from the SM
\cite{FM-BpsiK}.

Obviously, because of the $\phi_d^{\rm NP}$ term in (\ref{phid-NP}),
we may not convert the experimental information on $\phi_d$
into a constraint on $\beta$ in the presence of NP contributions to 
$B^0_{d,s}$--$\overline{B^0_{d,s}}$ mixing. Moreover, we may 
not use the SM interpretation of the $B_{d,s}$ mass differences
$\Delta M_{d,s}$ to determine the unitarity-triangle side
\begin{equation}\label{Rt-def}
R_t\equiv\left|\frac{V_{td}V_{tb}^\ast}{V_{cd}V_{cb}^\ast}\right|
=\frac{1}{\lambda}\left|\frac{V_{td}}{V_{cb}}\right|=
\sqrt{(1-\overline{\rho})^2+\overline{\eta}^2},
\end{equation}
as recently discussed in \cite{RF-PHYS-REP,rising,Becirevic}.
On the contrary, the determination of the side
\begin{equation}\label{Rb-def}
R_b\equiv\left|\frac{V_{ud}V_{ub}^\ast}{V_{cd}V_{cb}^\ast}\right|
=\left(1-\frac{\lambda^2}{2}\right)\frac{1}{\lambda}\left|
\frac{V_{ub}}{V_{cb}}\right|=\sqrt{\overline{\rho}^2+\overline{\eta}^2}
\end{equation}
by means of exclusive and inclusive transitions of the
type $b\to u \ell\overline{\nu}_\ell$ and
$b\to c\ell\overline{\nu}_\ell$,
which are dominated by SM tree-level amplitudes,
is very robust as far as the impact of NP is concerned \cite{GNW}.

A determination of the ``true'' unitarity triangle in the presence of a
completely general NP model, i.e.\ with arbitrary flavour-mixing
terms, is almost impossible. However, we may still perform useful
predictive analyses within certain scenarios for physics beyond the
SM, for example within models with ``minimal flavour violation''
(MFV), where the only source for flavour mixing is still given by the
CKM matrix (see, for instance, \cite{Univ_tri,DGIS}). In the present
paper we go one step beyond the MFV models: we analyse a scenario
with large generic NP contributions to $B^0_d$--$\overline{B^0_d}$
mixing ($\Delta B=2$), eventually also to $\varepsilon_K$
($\Delta S=2$), but {\it not} to the $\Delta B=1$ and $\Delta S=1$ decay
amplitudes. As we shall demonstrate, this scenario is well motivated by simple
dimensional arguments in a wide class of models,
including supersymmetric frameworks.

If we assume NP of this kind, we may complement the experimental
information on $\phi_d$ with data on CP violation in $B_d\to\pi^+\pi^-$
to extract $\gamma$, and may then fix the ``true'' apex of the unitarity
triangle with the help of the side $R_b$. Needless to note, we may then
also extract $\alpha$ and $\beta$. Several years ago, assuming also such 
a NP scenario but neglecting penguin effects, it was already pointed out 
in \cite{GNW} that it is actually possible to determine the unitarity 
triangle -- up to discrete ambiguities -- by combining the CP-violating 
observables of $B_d\to J/\psi K_{\rm S}$ and $B_d\to\pi^+\pi^-$ decays with  
$R_b$. Since we have now strong experimental and theoretical indications 
for large penguin effects in $B_d\to\pi^+\pi^-$, we must definitely care 
about them. To this end, we follow \cite{RF-Bpipi,FlMa2}, and use data 
on the CP-averaged $B_d\to\pi^\mp K^\pm$ branching ratio, which allows 
us to control the penguin contributions with the help of plausible 
dynamical assumptions and $U$-spin flavour-symmetry arguments. Employing
``QCD factorization'' to deal with the penguins topologies \cite{BBNS3}, 
a similar analysis was also performed in \cite{neubert}. Further alternative 
strategies to extract physics information from $\phi_d$, $R_b$, 
and CP violation in $B_d\to\pi^+\pi^-$ can be found, for instance, in 
\cite{alt-strat}.

Another important aspect of our analysis are the very rare decays
$K\to\pi\nu\overline{\nu}$ and $B_{d,s}\to\mu^+\mu^-$. Using the
allowed ranges for the generalized Wolfenstein parameters
obtained from the strategy sketched above, we are able to predict
the branching ratios for these modes independently of possible
NP contributions to $B^0_d$--$\overline{B^0_d}$ mixing. As we
shall show, these predictions could be very different from the SM
expectations.

The outline of this paper is as follows: a closer look at 
$\phi_d$, $R_b$ and $\beta$ is presented in Section~\ref{sec:phid}. 
In Section~\ref{sec:NP}, we motivate the scenario
where NP manifests itself only through contributions to
$B^0_d$--$\overline{B^0_d}$ mixing. In Section~\ref{sec:UT-det}, we discuss
how we may determine the ``true'' unitarity triangle for this kind
of NP with the help of CP violation in $B_d\to\pi^+\pi^-$.
The implications for the rare decays $K\to\pi\nu\overline{\nu}$
and $B_{d,s}\to\mu^+\mu^-$ are presented in Section~\ref{sec:rare}.
Finally, we conclude in Section~\ref{sec:concl}
with a short summary of our results and a brief outlook.

\boldmath
\section{A Closer Look at $\phi_d$, $R_b$ and $\beta$}\label{sec:phid}
\unboldmath
The present world average for $\sin\phi_d$, determined from mixing-induced
CP violation in $B_d\to J/\psi K_{\rm S}$ and similar modes, is
$\sin\phi_d=0.734\pm0.054$ \cite{nir}, which implies
\begin{equation}\label{phid-exp}
\phi_d=\left(47^{+5}_{-4}\right)^\circ \, \lor \,
\left(133^{+4}_{-5}\right)^\circ.
\end{equation}
Here the former solution would be in perfect agreement with the ``indirect''
range implied by the CKM fits, $40^\circ\lsim\phi_d\lsim60^\circ$
\cite{CKM-fits}, whereas the latter would show a significant discrepancy.
Measuring the sign of $\cos\phi_d$, both solutions can be distinguished.
There are several strategies on the market to accomplish this important task
\cite{ambig}. For example, in the $B\to J/\psi K$ system,
$\mbox{sgn}(\cos\phi_d)$ can be extracted from the time-dependent angular
distribution of the decay products of
$B_d\to J/\psi[\to\ell^+\ell^-] K^\ast[\to\pi^0K_{\rm S}]$,
if the sign of a hadronic parameter $\cos\delta_f$, involving a strong
phase $\delta_f$, is fixed through factorization \cite{DDF2DFN}. This
analysis is already in progress at the $B$ factories \cite{itoh}.

In this context, it is important to note that the CKM factor $R_b$
introduced in (\ref{Rb-def}) allows us to obtain the following bounds
(see Fig.~\ref{fig:UT}a):
\begin{equation}
(\sin\beta)_{\rm max}=R_b^{\rm max}, \quad
(\sin2\beta)_{\rm max}=2R_b^{\rm max}\sqrt{1-(R_b^{\rm max})^2}.
\end{equation}
Using the rather conservative experimental range
\begin{equation}\label{Rb-range}
R_b=0.38\pm0.08,
\end{equation}
which corresponds to $R_b^{\rm max}=0.46$, we obtain
\begin{equation}\label{beta-bound}
|\beta|_{\rm max}=27.4^\circ.
\end{equation}
Since the determination of $R_b$ from semileptonic $B$ decays, which
originate from tree-level SM processes, appears
to be very robust as far as the impact of NP is concerned, we
may consider (\ref{beta-bound}) as an upper bound for the ``true'' angle
$\beta$ of the unitarity triangle. Consequently, as we have $\phi_d=2\beta$
within the SM, (\ref{beta-bound}) implies
\begin{equation}\label{phid-SM-bound}
|\phi_d^{\rm SM}|_{\rm max}=2|\beta|_{\rm max}=55^\circ.
\end{equation}
Whereas the former solution $\phi_d\sim 47^\circ$ in (\ref{phid-exp})
satisfies this bound nicely, this is definitely not the case for
$\phi_d\sim 133^\circ$. The latter solution cannot be accommodated
in the SM and requires NP contributions to
$B^0_d$--$\overline{B^0_d}$ mixing, i.e.\ we now need 
$\phi_d^{\rm NP}\not=0$ in (\ref{phid-NP}). At this point, the important
question of how we may represent this second solution in the
$\overline{\rho}$--$\overline{\eta}$ plane arises. In the literature,
this is usually simply done through a second branch, corresponding to
\begin{equation}\label{wrong-branch}
\beta\sim133^\circ/2=66.5^\circ \quad \mbox{and} \quad
\beta\sim-(180^\circ-66.5^\circ)=-113.5^\circ,
\end{equation}
as we have shown in Fig.\ \ref{fig:UT}(b). However, because of
(\ref{phid-NP}), this is actually {\it not} correct. Moreover, since
$\phi_d\sim 133^\circ$ is associated with NP contributions to
$B^0_d$--$\overline{B^0_d}$ mixing, we may {\it no longer} use the
SM interpretation of $\Delta M_d$ and $\Delta M_s/\Delta M_d$
to determine the side $R_t$ of the unitarity triangle. In particular,
we may no longer conclude from the experimental lower bound on $\Delta M_s$
that $|\gamma|<90^\circ$. It is important to note
that (\ref{phid-NP}) applies to
$\phi_d\sim47^\circ$ as well, so that NP may -- at least in
principle -- hide itself in this case, despite the striking consistency
with the SM interpretation of other observables.

As already pointed out, one of the key ingredients to
determine the ``true'' apex of the unitarity triangle
without using information on $B^0_d$--$\overline{B^0_d}$ mixing
is the side $R_b$. If we complement it with a measurement
of $\gamma$, we may determine the coordinates of the apex of the unitarity
triangle straightforwardly through
\begin{equation}\label{UT-fix}
\overline{\rho}=R_b\cos\gamma, \quad \overline{\eta}=R_b\sin\gamma.
\end{equation}From a
theoretical point of view, certain pure ``tree'' decays would
be best suited for the determination of $\gamma$, allowing theoretically
clean extractions of this angle that would also be very robust
under the impact of NP (see, for instance,
\cite{tree-gam}).\footnote{In principle, NP could enter in
these strategies through $D^0$--$\overline{D^0}$ mixing. However, it
could then be taken into account through a measurement of the corresponding
mixing parameters \cite{D-mix-incl}.} A similar comment applies to the new
strategies that were recently proposed in \cite{RF-BDfr1,RF-BDfr2}. Here
decays of the kind $B_d\to D K_{\rm S (L)}$ and
$B_s\to D \eta^{(')}, D\phi, ...$ are employed. If we use the
$B^0_q$--$\overline{B^0_q}$ mixing phase $\phi_q$ as an input, and observe
the neutral $D$ mesons through their decays into CP-even and CP-odd
eigenstates, certain ``untagged'' and mixing-induced observables allow a
very efficient, theoretically clean determination of $\gamma$ in an
essentially unambiguous manner \cite{RF-BDfr1}; alternatively, we may also
employ decays of the neutral $D$ mesons into CP non-eigenstates to
this end \cite{RF-BDfr2}. These strategies appear to be
particularly interesting for the next generation of dedicated $B$
experiments, LHCb and BTeV, as well as those at super-$B$ factories,
although important steps may already be made at BaBar and Belle.

Since we cannot yet confront these methods with data, we have to employ
alternative strategies to extract $\gamma$. In this context,
$B_d\to\pi^+\pi^-$ offers a very interesting tool. If we use $\phi_d$ as an
input, and employ the CP-averaged $B_d\to\pi^\mp K^\pm$ branching
ratio to control the penguin effects \cite{RF-Bpipi}, we may determine
$\gamma$ from the CP-violating $B_d\to\pi^+\pi^-$ observables for each
of the two solutions given in (\ref{phid-exp}) \cite{FlMa2}; using
``QCD factorization'' \cite{BBNS3} to deal with the penguin effects, 
such an exercise was also performed in \cite{neubert}. In the
future, the $B_d\to\pi^+\pi^-$ observables can be fully exploited
with the help of $B_s\to K^+K^-$ \cite{RF-BsKK}. In contrast to the pure
tree decays mentioned above, these modes may in principle well be affected 
by new physics, as they receive contributions from loop-induced
flavour-changing neutral-current (FCNC) amplitudes.
However, as we shall discuss in the next section,
there is a particularly interesting
class of extensions of the  SM where new
physics yields sizeable contributions to $B^0_d$--$\overline{B^0_d}$
mixing only, leaving the decay amplitudes -- including the loop-induced
ones -- almost unaffected.

\boldmath
\section{New Physics in $B^0_d$--$\overline{B^0_d}$ Mixing}\label{sec:NP}
\unboldmath
\subsection{General Considerations}\label{sect:GC}
As far as flavour physics is concerned, extensions of the SM can be classified
into two wide categories: models with minimal flavour violation and models
with new sources of flavour mixing. Within the highly-constrained class of 
MFV models, the only source of flavour-sym\-metry breaking terms is given
by the SM Yukawa couplings \cite{DGIS}. As a consequence, all flavour-changing
transitions are still ruled by the CKM matrix. For this reason, many 
of the standard CKM constraints hold also in MFV models:
the determination of $R_t$ in terms of
$\Delta M_s/\Delta M_d$ is still valid~\cite{Univ_tri},
and the relation between $\sin\phi_d$ and $\sin 2\beta$ can differ
at most by an overall sign~\cite{BF2001}, i.e.\ $\phi_d^{\rm NP}=0^\circ$ 
or $\phi_d^{\rm NP}=180^\circ$. It is then easy to realize that within 
MFV models the structure of the unitarity triangle is fixed -- up to a 
twofold ambiguity -- even without $B_d\to\pi^+\pi^-$ data, and that there 
is no room for the  non-standard solution $\phi_d \sim 133^\circ$.

In the wide class of models with new sources
of flavour mixing, it is rather natural to
expect extra $\cO(1)$ contributions
to the \BBbar amplitude, with arbitrary phases, so that we
may accommodate any value of $\phi_d^{\rm NP}$.
It is also very natural to assume that these NP effects
have a negligible impact on $\Delta B=1$ amplitudes dominated by 
tree-level SM contributions; this happens essentially 
in all realistic models. On the other hand, it is less obvious why these
new sources of flavour-symmetry breaking
-- being able to induce $\cO(1)$  corrections to \BBbar mixing --
should have a small impact on $\Delta B=1$
amplitudes arising only at the loop level
within the SM. This hypothesis, which is one of
the main assumptions of the present analysis,
certainly does not represent the most general
NP scenario. However, as we shall discuss in the following, it can be 
realized under rather general conditions. 

\medskip

The generic NP scenario we shall advocate is a model where 
non-standard effects are negligible in all amplitudes 
that receive tree-level SM contributions (independently of 
possible CKM suppressions). Moreover, in order to protect 
the effects on $\Delta B=1$  FCNC amplitudes, 
we shall add the following two general requirements:
\begin{enumerate}
\item[{\bf i)}] the effective scale of NP is substantially
higher than the electroweak scale;
\item[{\bf ii)}] the adimensional effective couplings
ruling $\Delta B=2$ transitions can always be expressed
as the square of two $\Delta B=1$ effective couplings.
\end{enumerate}
Employing an effective-theory language, what we mean
under these two hypotheses is that the generic dimen\-sion-six
operators encoding NP contributions to
\BBbar and ${\Delta B=1}$ transitions can be written as
\beq
Q^{\rm NP}_{\Delta B=2} = \frac{ \delta_{bd}^2 }{\Lambda_{\rm eff}^2}
(\overline{b} \Gamma d) (\overline{b} \Gamma d),
\qquad
Q^{\rm NP}_{\Delta B=1} = \frac{ \delta_{bd}   }{\Lambda_{\rm eff}^2}
(\overline{b} \Gamma d) (\overline{q} \Gamma q),
\label{eq:NPops}
\eeq
where $\delta_{bd}$ denotes the new  $\Delta B=1$ effective
flavour-changing coupling, $\Gamma$ indicates generic
Dirac and/or colour structures, and possible coefficient
functions of $\cO(1)$ have been ignored. 
Coherently with the requirement ii), 
we shall also assume that $({\rm SU}(2)_{\rm L} \times  
{\rm U}(1)_{\rm Y})$-breaking operators of dimension less than six, 
such as the chromomagnetic operator, play a negligible role. The overall
normalization of the two operators in (\ref{eq:NPops}),
or the definition of the effective scale $\Lambda_{\rm eff}$,
has been chosen such that the corresponding SM $\Delta B=2$ term is
\beq
Q^{\rm SM}_{\Delta B=2} = \frac{ (V_{tb}^* V_{td})^2 }{ M_W^2}
(\overline{b} \Gamma d) (\overline{b} \Gamma d).
\label{eq:normal2}
\eeq
Choosing this normalization, we have implicitly factorized out
an overall coefficient of $\cO[(g/\sqrt{2})^4/(16\pi^2)]$ in the
effective Hamiltonian, both in the $\Delta B=2$ and in the 
$\Delta B=1$ cases. As a result, within the SM,
the loop-induced $\Delta B=1$ operators
can be written as
\beq
Q^{\rm SM}_{\Delta B=1} = {\cal C} \frac{ V_{tb}^* V_{td}  }{ M_W^2}
(\overline{b} \Gamma d) (\overline{q} \Gamma q),
\label{eq:normal1}
\eeq
where the coefficient function ${\cal C}$ is of $\cO(1)$
in the case of pure short-distance-dominated 
electroweak operators (such as those generated 
by $Z$-penguin and $W$-box diagrams)
and substantially larger than 
unity for those that receive large 
logarithmic corrections via RGE. 

Note that the normalization of the NP contributions in 
(\ref{eq:normal2}) does not necessarily imply that
they are generated only by loop amplitudes
at the fundamental level: their natural size 
is comparable to that of SM loop amplitudes, 
but they could well be induced by tree-level 
processes. For instance, a model 
with a heavy $Z'$ boson with a FCNC coupling to the 
$\overline{b} \Gamma d$ current would also perfectly 
fit in this scheme, and in this case {\em both}
$Q^{\rm SM}_{\Delta B=1}$ and $Q^{\rm SM}_{\Delta B=2}$
would receive tree-level contributions.
What we cannot  accommodate in this scheme is
the possibility that $Q^{\rm SM}_{\Delta B=1}$
receives large tree-level contributions 
and $Q^{\rm SM}_{\Delta B=2}$ only loop-induced ones: 
in our language, this case would correspond 
to a violation of the condition ii).

Since the measurement of the $B^0_d$--$\overline{B^0_d}$ mass difference 
falls in the ballpark of the SM expectations, the new flavour-changing 
coupling $\delta_{bd}$ cannot be arbitrarily large: barring fine-tuned 
scenarios with severe cancellations among different terms, we can allow 
at most $\cO(1)$ corrections to the SM amplitude. This implies (see also
\cite{FM-BpsiK})
\beq
\frac{\langle Q^{\rm NP}_{\Delta B=2} \rangle}{\langle
Q^{\rm SM}_{\Delta B=2} \rangle}
\lsim 1         \qquad \rightarrow \qquad
\frac{ \delta_{bd}}{  \Lambda_{\rm eff}}  \lsim
\frac{ V_{tb}^* V_{td} }{ M_W },
\label{eq:cond}
\eeq
where possible $\cO(1)$ factors associated with the matrix elements
of the operators have been neglected.

Owing to the different parametric dependence from scale
factor and flavour-changing coupling of $\Delta B=2$ and  $\Delta B=1$
operators in (\ref{eq:NPops}), if the condition (\ref{eq:cond}) is 
fulfilled the corresponding non-standard effects induced in $\Delta B=1$
flavour-changing neutral-current transitions turn out to be suppressed 
at least by a factor $\cO(M_W/\Lambda_{\rm eff})$
relative to the SM level: 
\beq
\frac{\langle Q^{\rm NP}_{\Delta B=2}
\rangle}{\langle Q^{\rm SM}_{\Delta B=2} \rangle}
\lsim 1         \qquad \rightarrow \qquad
\frac{\langle Q^{\rm NP}_{\Delta B=1} \rangle}{\langle
Q^{\rm SM}_{\Delta B=1} \rangle}
\leq  \frac{1}{{\cal C}} \frac{M_W}{\Lambda_{\rm eff}}.
\label{eq:cond2}
\eeq
Since the coefficient ${\cal C}$ is substantially larger than 1 for 
QCD-penguin amplitudes, the suppression is even more severe in this case.
We have thus obtained a natural justification for the smallness of 
non-standard effects in $\Delta B=1$ loop-induced amplitudes in the 
well-motivated scenario of a heavy NP scale ($\Lambda_{\rm eff} \gg M_W$).

\medskip 

Although rather qualitative, the above argument has the great advantage
of being almost independent of the details of the NP model.
Indeed, it can be realized in very different frameworks, from
low-energy supersymmetry to models with large extra dimensions.
As can be easily understood, this argument does not apply only to
$B^0_d$--$\overline{B^0_d}$ mixing:
it is characteristic of any type of $\Delta F=2$
meson--antimeson mixing versus the corresponding
$\Delta F=1$ loop-induced amplitudes, provided the corresponding assumption 
ii) is fulfilled. For this reason, in the phenomenological determination
of the CKM matrix of Section~\ref{sec:UT-det},
we shall try to avoid the use of observables such as $\varepsilon_K$
or $\Delta M_s$, which are sensitive to the $K^0$--$\overline{K^0}$
and $B^0_s$--$\overline{B^0_s}$ mixing amplitudes, respectively.

In principle, the case of $b\to s$ transitions is somehow 
different from the $b\to d$ and  $s\to d$ ones, since the 
$B^0_s$--$\overline{B^0_s}$ mass difference has 
not yet been measured. This is indeed one of the reasons 
why speculations about possible large NP effects in 
penguin-mediated $b\to s$ transitions, 
such as $B\to \pi K$ and -- especially $B\to \phi K$ -- are still 
very popular (see, for instance, \cite{BpiK-NP} and references therein). 
However, we recall that the available data on $\Delta M_s$ already show 
a preference for this observable to be close to its SM expectation 
\cite{CKM-fits}. If we assume that NP effects in  $\Delta M_s$
can be at most of $\cO(1)$, we can accommodate 
large NP effects in $b\to s$ transitions
only by means of violations of the conditions 
i) and ii) \cite{FM-BphiK}, or by fine-tuning. Therefore, in order to 
understand the consistency of our scenario, it will be very interesting 
to follow the evolution of future measurements 
of $B\to \pi K $ and $B\to \phi K$ decays. So far, the available data
on the $B\to \pi K$ observables fall well into the SM-allowed regions in
observable space \cite{FlMa2}. In particular, they do not indicate any
anomalous behaviour of the $B_d\to\pi^\mp K^\pm$ mode, which will be used
in Section~\ref{sec:UT-det} to control the penguin effects in 
$B_d\to\pi^+\pi^-$.

To conclude this general discussion, it is worth stressing 
that this mechanism is not representative of all possible models
with new sources of flavour mixing.
As already mentioned, the condition ii) is not necessarily
fulfilled. Moreover, a large hierarchy of matrix elements could
invalidate the conditions (\ref{eq:cond}) and (\ref{eq:cond2}).
This happens, for instance, in supersymmetry with specific choices 
of the soft-breaking terms. However, as we shall discuss in the 
following, it is very natural to assume that this mechanism works
also within supersymmetric models, so that the largest NP effects 
appear only in $\Delta F=2$ amplitudes.

\subsection{The Supersymmetry Case}
Among specific extensions of the SM, low-energy supersymmetry
is certainly one of the most interesting and well-motivated
possibilities. In the absence of a MFV pattern for the soft-breaking terms,
sizeable modifications of FCNC amplitudes are naturally expected
within this framework, and the mass-insertion approximation
provides a very efficient tool to describe them~\cite{GGMS96}.

Supersymmetric contributions to \BBbar mixing have been widely
discussed in the literature (see \cite{Becirevic} and references therein),
and there exist several possibilities to accommodate arbitrary values
of $\phi_d^{\rm NP}$.
For instance, we can simply adjust the coupling $\delta^D_{b_{\rm L} 
d_{\rm L}}$ (or, equivalently,  $\delta^D_{b_{\rm R}d_{\rm R}}$) to 
produce the desired modification of $B_d^0$--$\overline{B_d^0}$ mixing
via gluino-mediated box diagrams. In this case
the sum of supersymmetric and SM
contributions to \BBbar mixing can be written as
\beq
  \Delta M_d e^{-i\phi_d}
&\propto&  \frac{G^2_{\rm F} M^2_W }{2 \pi^2 } \eta_B  S_0(x_{t})
       \langle \overline{B_d^0} | \left(\overline{b}_{\rm L} 
\gamma_\mu d_{\rm L}\right)^2
| B_d^0 \rangle \no  \\
  && \times \left\{ \left[ V_{tb}^{*}V_{td} \right]^2  +
    \frac{\alpha_{\rm S}({\tilde M}_q)^2 \sin^4\Theta_W
M_W^2 }{ \alpha_{\rm e.m.}(M_Z)^2 {\tilde M}_q^2 }
     r_\eta \frac{ F_0(x_{qg}) }{  S_0(x_{t}) }
\left[ \delta^D_{b_{\rm L} d_{\rm L}} \right]^2 \right\},
\label{eq:BB_SUSY}
\eeq
where $S_0( x_{t}\equiv m_t^2/M_W^2 )$ and $\eta_B$ denote the initial 
condition and leading QCD corrections of the SM Wilson coefficient,
respectively (see, e.g., \cite{BBL}). The explicit expression for $F_0$ -- 
the supersymmetric loop function depending on the ratio 
$x_{qg}\equiv{\tilde M_q}^2/{\tilde M_g}^2$ of squark and gluino
masses -- and the corresponding 
QCD correction factor $r_\eta$ can be found in \cite{Becirevic}. As a 
reference figure, note that $F_0(1)/S_0(x_{t})=1/27/S_0(x_{t})\approx 0.015$, 
and that we can set, to a good approximation, $r_\eta \approx 1$.

Expression (\ref{eq:BB_SUSY}) provides a simple realization of
the general scenario discussed in the previous subsection,
with the adimensional effective flavour-changing coupling given by 
$\delta^D_{b_{\rm L}d_{\rm L}}$, and the effective NP scale given by
\beq
 \Lambda^{\rm SUSY}_{\rm eff} =  \frac{  \alpha_{\rm e.m.}(M_Z)
{\tilde M}_q }{
    \alpha_{\rm S}({\tilde M}_q) \sin^2\Theta_W  }
    \left| \frac{ S_0(x_{t}) }{ r_\eta F_0(x_{qg}) } \right|^{1/2}
    \approx 1.4~{\rm TeV} \times \left( \frac{ {\tilde M}_q }{ \rm 0.5~TeV}
\right).
\eeq
According to condition (\ref{eq:cond}),
corrections to \BBbar mixing of  $\cO(1)$ should be obtained for
$|\delta^D_{b_{\rm L} d_{\rm L}}| \sim 0.1  \times 
( {\tilde M}_q / 0.5~{\rm TeV})$: this expectation is fully 
confirmed by the detailed analysis of \cite{Becirevic}. As 
pointed out in \cite{rising},
since SM and supersymmetric loop functions have
the same sign,\footnote{This statement is not valid for
arbitrary $x_{qg}$; however, it holds for $x_{qg} \sim 1$,
which is suggested by RGE constraints in
grand-unified scenarios~\cite{RGEb}.}
$\delta^D_{b_{\rm L} d_{\rm L}}$ must have a large imaginary part
if the SM term gives too large a $\Delta M_{d}$,
as in the case where the ``true value'' of $\gamma$ lies in the
second quadrant. Interestingly, as we shall see in the following section,
this scenario is the favoured one for $\phi_d \sim 133^\circ$.

Owing to the general dimensional argument in (\ref{eq:cond2}),
a supersymmetric framework with ${\tilde M}_q \gsim  0.5~{\rm TeV}$, where 
$|\delta^D_{b_{\rm L} d_{\rm L}}| \sim 0.1  
\times ( {\tilde M}_q / 0.5~{\rm TeV})$
is the only new source of flavour-symmetry breaking,
leads to negligible effects in $\Delta B =1$ transitions.
This can be explicitly verified by means of \cite{Lunghi}
in the case of electroweak penguin amplitudes of the type
$b \to (s,d)  \ell^+  \ell^- $, or by means of
\cite{BRS} in the case of $s \to d  \nu  \bar{\nu}$ transitions.
According to these analyses, if ${\tilde M}_q \gsim  0.5~{\rm TeV}$
and if we have only left--left mass insertions,
the supersymmetric corrections
to electroweak penguin amplitudes reach at most the level
of a few per cent with respect to the SM contributions.
These results are particularly important for the discussion
of rare decays in Section~\ref{sec:rare}: they show that it is 
perfectly conceivable to have a scenario where the NP corrections to 
$B^0_d$--$\overline{B^0_d}$ mixing are large, but the direct NP 
contributions to the rare decay amplitudes are negligible.

Within this framework, i.e.\ with heavy
squark masses and non-standard sources of
flavour-symmetry breaking induced only
by left--left mass insertions,
the relative impact of non-standard effects
is even smaller for non-leptonic QCD-penguin
amplitudes, such as those contributing to
$B_d \to \pi^+\pi^-$ and $B_d\to\pi^\mp K^\pm$ decays. 
Indeed, in this case the SM contribution is substantially enhanced 
by the coupling constant and by large (infra-red)
logarithms~\cite{BS}.

On the other hand, 
as already mentioned in the general discussion,
we recall that this framework is not
representative of all possible supersymmetric scenarios.
For instance, it is well known that by means of
flavour-non-diagonal $A$ terms, i.e.\ mass insertions
of the left--right type, we can generate
sizeable effects in $\Delta F=1$ transitions --
in particular rare decays -- without contradicting the 
$\Delta F=2$ bounds \cite{CIMM}.
This happens because we may generate by means of left--right 
mass insertions sizeable contributions to
$({\rm SU}(2)_{\rm L} \times  {\rm U}(1)_{\rm Y})$-breaking
$\Delta F=1$ operators, which allow us to evade 
condition  ii) \cite{BCIRS}. Similarly, condition ii) is 
badly violated by Higgs-mediated FCNC amplitudes in large 
$\tan\beta$ scenarios (see, in particular, 
\cite{BK}). Also in this case the reason is intimately related to 
an interplay between flavour- and electroweak-symmetry breaking~\cite{DGIS}.

To summarize, we can state that the scenario with large non-standard 
effects only in $\Delta F=2$ amplitudes is characteristic of supersymmetric 
models with: i) a heavy scale for the soft-breaking terms, ii) 
new sources of flavour-symmetry breaking only (or mainly) in the 
soft-breaking terms which do not involve the Higgs fields, iii) 
Yukawa interactions very similar to the pure SM case.

\boldmath
\section{Determination of the Unitarity Triangle}\label{sec:UT-det}
\unboldmath
Let us now assume that we have NP of the kind
specified in the previous section. We may then complement the
experimentally determined $B^0_d$--$\overline{B^0_d}$ mixing phase
$\phi_d$ and the CKM factor $R_b$ with data on CP violation in the
$B$-factory benchmark mode $B_d\to\pi^+\pi^-$ to fix the apex of the
``true'' unitarity triangle in the $\overline{\rho}$--$\overline{\eta}$
plane.

\boldmath
\subsection{CP Violation in $B_d\to\pi^+\pi^-$}
\unboldmath
The decay $B_d^0\to\pi^+\pi^-$ originates from
$\overline{b}\to\overline{u}u\overline{d}$ quark-level transitions.
Within the SM and the scenario for NP introduced
in Section~\ref{sec:NP}, we may write the corresponding decay amplitude
as follows \cite{RF-BsKK}:
\begin{equation}\label{Bpipi-ampl}
A(B_d^0\to\pi^+\pi^-)\propto\left[e^{i\gamma}-de^{i\theta}\right],
\end{equation}
where the CP-conserving strong parameter $d e^{i\theta}$ measures -- sloppily
speaking -- the ratio of penguin to tree contributions in $B_d\to\pi^+\pi^-$.
If we had negligible penguin contributions, i.e.\ $d=0$, the corresponding
CP-violating observables were simply given by
\begin{equation}\label{Bpipi-CP0}
{\cal A}_{\rm CP}^{\rm dir}(B_d\to\pi^+\pi^-)=0, \qquad
{\cal A}_{\rm CP}^{\rm mix}(B_d\to\pi^+\pi^-)=\sin(\phi_d+2\gamma)  
~{\stackrel{\rm SM }{=}} -\sin 2\alpha,
\end{equation}
where we have, in the last identity, also used the SM relation 
$\phi_d=2\beta$ and the unitarity relation $2\beta+2\gamma=2\pi-2\alpha$.
We observe that actually the phases $\phi_d$ and $\gamma$
enter directly in the $B_d\to\pi^+\pi^-$ observables, and not $\alpha$. 
Consequently, since $\phi_d$ can be fixed straightforwardly through 
$B_d\to J/\psi K_{\rm S}$, we may use $B_d\to\pi^+\pi^-$ to probe $\gamma$, 
which has important advantages when dealing with penguin and NP effects 
\cite{RF-Bpipi,FlMa2,RF-BsKK}. This procedure was also adopted in the 
``QCD factorization'' analyses performed in \cite{BBNS3,neubert}. We shall 
come back to this point below.

Measurements of the CP-violating $B_d\to\pi^+\pi^-$ observables are
already available:
\begin{equation}\label{Adir-exp}
{\cal A}_{\rm CP}^{\rm dir}(B_d\to\pi^+\pi^-)=\left\{
\begin{array}{ll}
-0.30\pm0.25\pm0.04 & \mbox{(BaBar \cite{BaBar-Bpipi})}\\
-0.77\pm0.27\pm0.08 & \mbox{(Belle \cite{Belle-Bpipi})}
\end{array}
\right.
\end{equation}
\begin{equation}\label{Amix-exp}
{\cal A}_{\rm CP}^{\rm mix}(B_d\to\pi^+\pi^-)=\left\{
\begin{array}{ll}
-0.02\pm0.34\pm0.05& \mbox{(BaBar \cite{BaBar-Bpipi})}\\
+1.23\pm0.41 ^{+0.07}_{-0.08} & \mbox{(Belle \cite{Belle-Bpipi}).}
\end{array}
\right.
\end{equation}
The BaBar and Belle results are unfortunately not fully consistent with
each other. Hopefully, the experimental picture will be clarified soon.
If we nevertheless form the weighted averages of (\ref{Adir-exp}) and
(\ref{Amix-exp}), using the rules of the Particle Data Group (PDG)
\cite{PDG}, we obtain
\begin{eqnarray}
{\cal A}_{\rm CP}^{\rm dir}(B_d\to\pi^+\pi^-)&=&-0.51\pm0.19 \,\, (0.23)
\label{Bpipi-CP-averages}\\
{\cal A}_{\rm CP}^{\rm mix}(B_d\to\pi^+\pi^-)&=&+0.49\pm0.27 \,\, (0.61),
\label{Bpipi-CP-averages2}
\end{eqnarray}
where the errors in brackets are the ones increased by the PDG
scaling-factor procedure. Direct CP violation at this level
would require large penguin contributions with large CP-conserving
strong phases. Interestingly, a significant impact of penguins
on $B_d\to\pi^+\pi^-$ is also indicated by data on the $B\to\pi K,\pi\pi$
branching ratios \cite{RF-Bpipi,FlMa2,GR-Bpipi}, as well as by theoretical
considerations \cite{BBNS3,PQCD-appl}. Consequently, it is already
evident that the penguin contributions to $B_d\to\pi^+\pi^-$ {\it cannot}
be neglected.

\boldmath
\subsection{Complementing $B_d\to\pi^+\pi^-$ with $B_d\to\pi^\mp K^\pm$}
\unboldmath
Over the recent years, many approaches to control the impact of the
penguin contributions on the extraction of weak phases from the
CP-violating $B_d\to\pi^+\pi^-$ observables were proposed (see,
for example, \cite{GR-Bpipi,BBNS3,Bpipi-strategies}). Let us here follow
the method suggested in \cite{RF-Bpipi,FlMa2}, which is a variant
of the $B_d\to\pi^+\pi^-$, $B_s\to K^+K^-$ strategy proposed in
\cite{RF-BsKK}. If we apply (\ref{Bpipi-ampl}), we may write (for
explicit expressions, see \cite{RF-BsKK}):
\begin{equation}\label{CP-Bpipi-par}
{\cal A}_{\rm CP}^{\rm dir}(B_d\to\pi^+\pi^-)=\mbox{fct}(d,\theta,\gamma),
\quad {\cal A}_{\rm CP}^{\rm mix}(B_d\to\pi^+\pi^-)=
\mbox{fct}(d,\theta,\gamma,\phi_d),
\end{equation}
which are {\it exact} parametrizations within the SM, and hold
also for the NP scenario specified in Section~\ref{sec:NP}. If we
fix $\phi_d$ through (\ref{phid-exp}), these two observables depend on
three unknown parameters, $d$, $\theta$ and $\gamma$. In order to extract
these quantities, it would be ideal to measure the following observables:
\begin{equation}
{\cal A}_{\rm CP}^{\rm dir}(B_s\to K^+K^-)=\mbox{fct}(d',\theta',\gamma),
\quad {\cal A}_{\rm CP}^{\rm mix}(B_s\to K^+K^-)=
\mbox{fct}(d',\theta',\gamma,\phi_s),
\end{equation}
where $\phi_s$ can be assumed to be negligibly small in the SM,
or can be fixed through CP-violating effects in $B_s\to J/\psi \phi$.
Since $B_s\to K^+K^-$ is related to $B_d\to\pi^+\pi^-$ through an
interchange of all strange and down quarks, i.e.\ through the $U$-spin
flavour symmetry of strong interactions, we may derive the $U$-spin
relation
\begin{equation}\label{U-spin-rel1}
d'=d,
\end{equation}
which allows us to determine $d$, $\theta$, $\theta'$ and $\gamma$ from
the CP-violating observables of the $B_s\to K^+K^-$, $B_d\to\pi^+\pi^-$
system \cite{RF-BsKK}. Unfortunately, $B_s\to K^+K^-$ is not accessible
at the $e^+e^-$ $B$ factories operating at the $\Upsilon(4S)$ resonance;
this decay can be observed for the first time at run II of the Tevatron
\cite{TEV-Report}, and can be ideally studied in the era of the LHC
\cite{LHC-Report}. However, since $B_s\to K^+K^-$ is related to
$B_d\to\pi^\mp K^\pm$ through an interchange of spectator quarks, we have
\begin{equation}\label{CP-rel}
{\cal A}_{\rm CP}^{\rm dir}(B_s\to K^+K^-)\approx 
{\cal A}_{\rm CP}^{\rm dir}(B_d\to \pi^\mp K^\pm)
\end{equation}
\begin{equation}\label{BR-rel}
\mbox{BR}(B_s\to K^+K^-)\approx\mbox{BR}(B_d\to \pi^\mp K^\pm)
\frac{\tau_{B_s}}{\tau_{B_d}},
\end{equation}
and may approximately use $B_d\to\pi^\mp K^\pm$, which has already been 
measured at the $B$ factories, to deal with the penguins in $B_d\to\pi^+\pi^-$
\cite{RF-Bpipi}. The key quantity is then the following ratio of the
CP-averaged $B_d\to\pi^+\pi^-$ and $B_d\to\pi^\mp K^\pm$ branching ratios:
\begin{equation}\label{H-det}
H\equiv\frac{1}{\epsilon}\left(\frac{f_K}{f_\pi}\right)^2
\left[\frac{\mbox{BR}(B_d\to\pi^+\pi^-)}{\mbox{BR}(B_d\to\pi^\mp K^\pm)}
\right]=
\left\{\begin{array}{cc}
7.4\pm2.5 & \mbox{(CLEO \cite{CLEO-BpiK})}\\
7.6\pm1.2 & \mbox{(BaBar \cite{BaBar-BpiK})}\\
7.1\pm1.9 & \mbox{(Belle \cite{Belle-BpiK})}
\end{array}\right\}=7.5\pm0.9,
\end{equation}
where the factor $f_K/f_\pi$ involving the kaon and pion decay constants
takes into account factorizable $U$-spin-breaking corrections, and
$\epsilon\equiv\lambda^2/(1-\lambda^2)$. If we employ -- in addition to
(\ref{U-spin-rel1}) -- another $U$-spin relation,
\begin{equation}\label{U-spin-rel2}
\theta'=\theta,
\end{equation}
and make plausible dynamical assumptions to replace the $B_s\to K^+K^-$
channel through $B_d\to\pi^\mp K^\pm$,\footnote{Note that
the $U$-spin relation (\ref{U-spin-rel1}) is sufficient in the case
of the $B_s\to K^+K^-$, $B_d\to\pi^+\pi^-$ strategy.} we may write
\begin{equation}\label{H-par}
H=\mbox{fct}(d,\theta,\gamma).
\end{equation}
Consequently, (\ref{CP-Bpipi-par}) and (\ref{H-par}) allow us now to
determine $\gamma$, as well as $d$ and $\theta$. The explicit formulae
to implement this strategy can be found in \cite{FlMa2}, taking also into
account possible corrections to (\ref{U-spin-rel1}) and (\ref{U-spin-rel2}).

As discussed in detail in \cite{FlMa2}, additional information is also 
provided by the direct CP asymmetry of the $B_d\to\pi^\mp K^\pm$ modes,
which is related to the one of $B_d\to\pi^+\pi^-$ through the  
following $U$-spin relation \cite{RF-Bpipi,RF-BsKK}:
\begin{equation}\label{U-spin-CP}
{\cal A}_{\rm CP}^{\rm dir}(B_d\to \pi^\mp K^\pm)\approx
{\cal A}_{\rm CP}^{\rm dir}(B_s\to K^+K^-)=-\epsilon H 
{\cal A}_{\rm CP}^{\rm dir}(B_d\to \pi^+\pi^-).
\end{equation}
If we insert $H\sim7.5$ and 
${\cal A}_{\rm CP}^{\rm dir}(B_d\to \pi^+\pi^-)\sim-0.5$ into this 
expression, we obtain 
${\cal A}_{\rm CP}^{\rm dir}(B_d\to \pi^\mp K^\pm)\sim+0.2$. On
the other hand, the present average of the $B$-factory results for this 
CP asymmetry is given by 
${\cal A}_{\rm CP}^{\rm dir}(B_d\to \pi^\mp K^\pm)=+0.09\pm0.04$
\cite{tomura}. In comparison with the old result 
${\cal A}_{\rm CP}^{\rm dir}(B_d\to \pi^\mp K^\pm)=+0.05\pm0.06$ considered
in \cite{FlMa2}, this CP asymmetry has interestingly moved towards larger
values, and its sign is in accordance with the expectation following from
(\ref{U-spin-CP}). In view of the considerable experimental uncertainties, 
the feature that ${\cal A}_{\rm CP}^{\rm dir}(B_d\to \pi^\mp K^\pm)
=+0.09\pm0.04$ still seems to favour somewhat smaller values than 
$\sim0.2$ cannot be considered as a discrepancy. Moreover, (\ref{CP-rel})
and (\ref{BR-rel}) rely not only on the $U$-spin symmetry, but also 
on dynamical assumptions, as ``exchange'' and ``penguin annihilation''
topologies contribute to the $B_s\to K^+K^-$ channel, but not to 
$B_d\to \pi^\mp K^\pm$. Although such topologies are expected
to play a minor r\^ole, they may in principle be enhanced through large 
rescattering effects. Once the $B_s\to K^+K^-$ channel is measured, we 
have no longer to rely on this dynamical assumption, and may check how well 
(\ref{CP-rel}) and (\ref{BR-rel}) are actually satisfied. In particular 
a measurement of the CP-averaged
$B_s\to K^+K^-$ branching ratio, which will hopefully be soon available from
the CDF collaboration, would already be a very important step ahead, 
allowing a considerably more solid determination of $H$. 

As pointed out in \cite{FlMa2,FlMa1}, there is a transparent way to
characterize the $B_d\to\pi^+\pi^-$ mode in the space of its
CP-violating observables, allowing a simple comparison with the
experimental data. Let us focus here on the unitarity triangle and the
allowed region for its apex in the $\overline{\rho}$--$\overline{\eta}$
plane.

\subsection{Fixing the Apex of the Unitarity Triangle}
If we follow \cite{FlMa2} and employ the quantity $H$ to control the
penguin effects in $B_d\to\pi^+\pi^-$, we may convert the direct
and mixing-induced CP asymmetries of this decay into values of $\gamma$.
Using $H=7.5$ and the experimental averages given in (\ref{Bpipi-CP-averages})
and (\ref{Bpipi-CP-averages2}), we obtain
\begin{equation}\label{gam-res}
35^\circ\lsim\gamma\lsim79^\circ \, (\phi_d=47^\circ), \quad
101^\circ\lsim\gamma\lsim145^\circ \, (\phi_d=133^\circ).
\end{equation}
As pointed out in \cite{FlMa2}, these solutions are related to each other
through
\begin{equation}\label{sym1}
\phi_d\to 180^\circ-\phi_d, \quad \gamma\to 180^\circ-\gamma.
\end{equation}
Because of the unsatisfactory experimental situation concerning the CP 
asymmetries of $B_d\to\pi^+\pi^-$, the ranges in (\ref{gam-res}) should
mainly be considered as an illustration of how our strategy is working.
Indeed, in order to obtain (\ref{gam-res}), we have just used the ``ordinary''
errors in (\ref{Bpipi-CP-averages}) and (\ref{Bpipi-CP-averages2}), and
not the enlarged ones given there in brackets. Interestingly, the
information on the {\it positive} sign of
${\cal A}_{\rm CP}^{\rm mix}(B_d\to\pi^+\pi^-)$, which is favoured by the
Belle measurement \cite{Belle-Bpipi}, already implies that $\gamma\sim60^\circ$
{\it cannot} be accommodated in the case of $\phi_d\sim 133^\circ$, as
emphasized in \cite{FlMa2}. The experimental uncertainties will be reduced
considerably in the future, thereby providing significantly more stringent
results for $\gamma$.

It should be noted that we have assumed in (\ref{gam-res}) --
as is usually done -- that $\gamma\in[0^\circ,180^\circ]$.
This range is implied by the interpretation of $\varepsilon_K$,
provided that: i) NP does not change the sign of the $\Delta S=2$
amplitude with respect to the SM; ii) the ``bag'' parameter $B_K$
is positive (as indicated by all existing non-perturbative calculations).
A similar assumption about the ``bag'' parameter $B_{B_q}$
-- the $B_q$-meson counterpart of $B_K$ --
enters also in (\ref{phid-NP}); for a discussion of the very
unlikely $B_K<0$, $B_{B_q}<0$ cases, see \cite{GKN}.
If we relaxed these assumptions about NP and/or ``bag''
parameters, we would need to double the solutions
and consider the specular case $\gamma\in[180^\circ,360^\circ]$.

\begin{figure}
\begin{center}
 \psfig{file=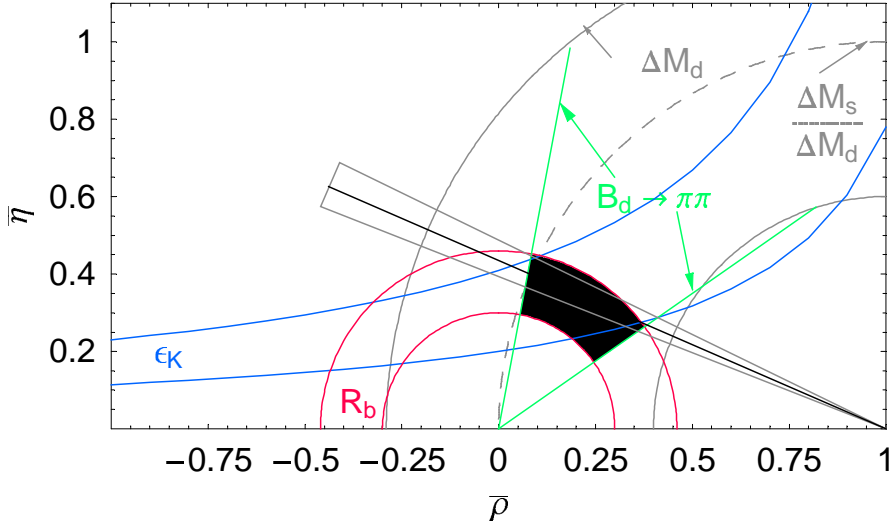,width=0.75\textwidth}
\end{center}
\caption{Allowed region for the apex of the unitarity triangle determined from 
the CP-violating $B_d\to\pi^+\pi^-$ observables given in
(\ref{Bpipi-CP-averages}) and (\ref{Bpipi-CP-averages2}) in the case of 
$\phi_d=47^\circ$ ($H=7.5$). For comparison, we also show the hyperbola 
corresponding to the SM interpretation of $\varepsilon_K$, the 
$\beta=23.5^\circ$ branch arising from the SM interpretation of 
$\phi_d$, the circle corresponding to the SM interpretation of 
$\Delta M_d$ (solid grey), and the upper bound on $R_t$ (dashed grey), 
which originates from the lower bound on $\Delta M_s$ 
(see Table~\ref{tab:inputs}).}\label{fig:first}
\end{figure}

\begin{figure}
\begin{center}
 \psfig{file=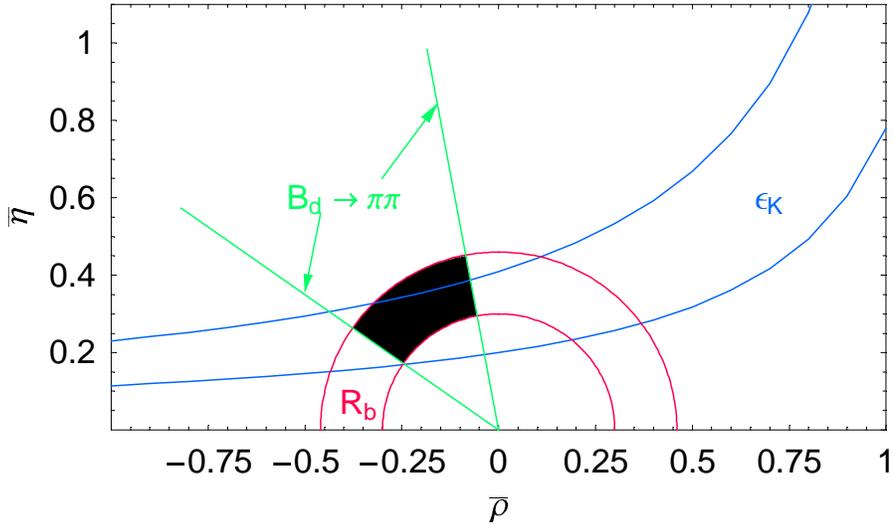,width=0.75\textwidth}
\end{center}
\caption{Allowed region for the apex of the unitarity triangle determined
from the CP-violating $B_d\to\pi^+\pi^-$ observables given in
(\ref{Bpipi-CP-averages}) and (\ref{Bpipi-CP-averages2}) in the case of
$\phi_d=133^\circ$ ($H=7.5$). The hyperbola corresponds to the
SM interpretation of $\varepsilon_K$.}
\label{fig:second}
\end{figure}

\begin{table}[t]
\begin{center}
\begin{tabular}{|l|l|} 
\hline
Experimental data & Hadronic parameters \\ \hline 
$\lambda =0.2241\pm0.0036$ &  ${\hat B}_{K}=0.85\pm0.15$ \\
$|V_{cb}| = 0.041 \pm 0.002$ & $f_{B_d}\sqrt{{\hat B}_{B_d}}=
(230 \pm 28 \pm 28) \, \mbox{MeV}$   \\
$R_b = 0.38 \pm 0.08$ & $f_{B_d}=(203 \pm 27 {}^{+0}_{-20}) \, \mbox{MeV}$\\
${\overline m_t}(m_t) = (167\pm 5) \, \mbox{GeV} $  &  
$f_{B_s}=(238 \pm 31) \, \mbox{MeV}$ \\
$\Delta M_d=(0.503\pm0.006) \, \mbox{ps}^{-1}$ & 
$\xi = 1.16$  \\ 
$\Delta M_s >14.4 \, \mbox{ps}^{-1}$ &  \\ 
\hline
\end{tabular}
\end{center}
\caption{Input values used to draw the $\varepsilon_K$,
$\Delta M_d$ and $\Delta M_s/\Delta M_d$  constraints 
in Figs.~\ref{fig:first} and \ref{fig:second}, and to analyse the branching 
ratios of rare decays in Section~\ref{sec:rare}. As usual, we define
$\xi\equiv(f_{B_s}/f_{B_d})\sqrt{{\hat B}_{B_s}/{\hat B}_{B_d}}$.}
\label{tab:inputs}
\end{table}

Using (\ref{gam-res}), we may now fix the apex of the
unitarity triangle. To this end, we insert (\ref{gam-res}) into
(\ref{UT-fix}), and use the range for $R_b$ given in (\ref{Rb-range}).
The results of this exercise are shown in Figs.~\ref{fig:first} and
\ref{fig:second} for $\phi_d=47^\circ$ and $\phi_d=133^\circ$, respectively.
Because of (\ref{sym1}), the values of $\overline{\rho}$ and
$\overline{\eta}$ corresponding to these two solutions for $\phi_d$ are
related to each other as follows:
\begin{equation}\label{sym2}
\overline{\rho}\to -\overline{\rho}, \quad \overline{\eta}\to\overline{\eta}.
\end{equation}
In order to guide the eye, we have also included in Figs.~\ref{fig:first} 
and \ref{fig:second} the well-known SM $\varepsilon_K$ hyperbola. In 
Fig.~\ref{fig:first}, we show also the $\beta=23.5^\circ$ branch, which 
corresponds to the SM interpretation of $\phi_d=47^\circ$, as well as the 
circle with radius $R_t$ around $(1,0)$ that is fixed through $\Delta M_d$, 
and the constraint arising from the lower bound on $\Delta M_s/\Delta M_d$. 
In order to draw the SM $\Delta M_{d,s}$ contours, as well as the one implied 
by $\varepsilon_K$,\footnote{We recall that these constraints are only shown 
for illustrative purposes, and are not used 
to fix the allowed region in the $\overline{\rho}$--$\overline{\eta}$ 
plane.} we use the parameters
collected in Table~\ref{tab:inputs}, which will also be employed in the
discussion of rare decays in Section~\ref{sec:rare}. It is remarkable that 
we obtain a perfect agreement of our $B_d\to\pi^+\pi^-$ range, which does 
{\it not} rely on $\varepsilon_K$ or $B^0_d$--$\overline{B^0_d}$ mixing, 
with {\it all} these constraints. 
On the other hand, the case of $\phi_d=133^\circ$ shown in
Fig.~\ref{fig:second} requires large NP contributions to
$B^0_d$--$\overline{B^0_d}$ mixing, so that we may there {\it no longer} 
use $\Delta M_d$ or $\Delta M_s/\Delta M_d$ to determine the side $R_t$ of 
the unitarity triangle, and may {\it no longer} convert $\phi_d$ directly 
into a straight line in the $\overline{\rho}$--$\overline{\eta}$ plane. For 
this reason, we have not included the corresponding contours in 
Fig.~\ref{fig:second}. Interestingly, our $\phi_d=133^\circ$ region would 
still be consistent with the $\varepsilon_K$ hyperbola, but would now 
favour values of $\gamma$ that are larger than $90^\circ$. The difference 
between the black range in this figure and the $\phi_d\sim133^\circ$ branch 
illustrated in Fig.\ \ref{fig:UT}(b), which is usually used in the literature 
to represent this case, is striking. As we have already emphasized, once we 
agree that the value of $R_b$ cannot be modified by NP effects, it is 
{\it not} correct to include this branch in the 
$\overline{\rho}$--$\overline{\eta}$ plane. 

Using the ``QCD factorization'' approach \cite{BBNS3}, a similar exercise 
was performed for the mixing-induced $B_d\to\pi^+\pi^-$ CP asymmetry in 
\cite{neubert}, i.e.\ this formalism was used to calculate the
hadronic parameter $de^{i\theta}$, and to extract $\gamma$ from 
(\ref{Bpipi-CP-averages2}) for the two solutions of $\phi_d$. However,
since ``QCD factorization'' points towards a small value of the direct
CP asymmetry ${\cal A}_{\rm CP}^{\rm dir}(B_d\to\pi^+\pi^-)\sim +0.1$ for 
$\gamma\in[0^\circ,180^\circ]$ \cite{BBNS3}, which would be in contrast 
to the large negative central value in (\ref{Bpipi-CP-averages}), it is 
unsatisfactory in our opinion to follow this interesting avenue to extract 
information on the unitarity trianlge from CP violation in 
$B_d\to\pi^+\pi^-$ for the time being. In our more phenomenological
approach, we are able to deal also with a large direct CP asymmetry in 
$B_d\to\pi^+\pi^-$, entering our range for $\gamma$. As we have already
noted, the present experimental situation has urgently to be clarified,
and it will be very exciting to see how the data will evolve. In the future,
we may well arrive at a picture that is consistent with the ``QCD 
factorization'' pattern. However, the present data leave also a lot of
space for other scenarios.

\begin{table}[t]
\vspace*{0.3truecm}
\begin{center}
\begin{tabular}{|c||c|c|c|c|c|}
\hline Case of $\tilde\xi=1$ & $\alpha$ & $\beta$ & $\gamma$  &
$\overline{\rho}$ &
$\overline{\eta}$\\
\hline $\phi_d=47^\circ$ & $(103\pm29)^\circ$ & $(20\pm7)^\circ$
&
$(57\pm22)^\circ$ & $+0.21\pm0.16$ & $+0.31\pm0.14$ \\
\hline $\phi_d=133^\circ$ & $(44\pm20)^\circ$ & $(15\pm7)^\circ$ &
$(123\pm22)^\circ$ & $-0.21\pm0.16$ & $+0.31\pm0.14$ \\
%\hline
%\end{tabular}
%
%\begin{tabular}{|c|c|c|c|c|c|}
\hline\hline
Case of $\tilde\xi=0.8$ & $\alpha$ & $\beta$ & $\gamma$  &
$\overline{\rho}$ &
$\overline{\eta}$\\
\hline $\phi_d=47^\circ$  &
$(102\pm33)^\circ$ & $(19\pm8)^\circ$ &
$(59\pm26)^\circ$ & $+0.20\pm0.18$ & $+0.31\pm0.15$ \\
\hline $\phi_d=133^\circ$  &
$(46\pm23)^\circ$ & $(16\pm8)^\circ$ &
$(121\pm26)^\circ$ & $-0.20\pm0.18$ & $+0.31\pm0.15$ \\
%\hline
%\end{tabular}
%
%\begin{tabular}{|c|c|c|c|c|c|}
\hline\hline Case of $\tilde\xi=1.2$ & $\alpha$ & $\beta$ &
$\gamma$  &
$\overline{\rho}$ & $\overline{\eta}$ \\
\hline  $\phi_d=47^\circ$ &
$(104\pm26)^\circ$ & $(20\pm7)^\circ$ &
$(56\pm19)^\circ$ & $+0.22\pm0.14$ & $+0.31\pm0.13$ \\
\hline  $\phi_d=133^\circ$ &
$(43\pm17)^\circ$ & $(15\pm7)^\circ$ &
$(124\pm19)^\circ$ & $-0.22\pm0.14$ & $+0.31\pm0.13$ \\
\hline
\end{tabular}
\caption{The ranges for the angles of the unitarity triangle and
the generalized Wolfenstein parameters corresponding to the black
regions shown
in Figs.~\ref{fig:first} and \ref{fig:second} ($\tilde\xi=1$).
In order to
illustrate the impact of possible $U$-spin breaking effects (see
(\ref{U-spin-break})), we give also the ranges arising for
$\tilde\xi=0.8$ and $\tilde\xi=1.2$. Note that the values for
$\overline{\rho}$ and $\overline{\eta}$ satisfy (\ref{sym2}).}\label{expval}
\end{center}
\vspace*{-0.3truecm}
\end{table}

The results shown in Figs.~\ref{fig:first} and \ref{fig:second} are
complemented by Table~\ref{expval}, where we collect the ranges for the
angles $\alpha$, $\beta$ and $\gamma$ of the unitarity
triangle, as well as those for the generalized Wolfenstein parameters
$\overline{\rho}$ and $\overline{\eta}$. In this table, we illustrate
also the impact of $U$-spin-breaking effects on (\ref{U-spin-rel1}),
which are described by the following parameter:
\begin{equation}\label{U-spin-break}
\tilde\xi=d'/d.
\end{equation}
As noted in \cite{FlMa2}, these effects are potentially much more important
than possible corrections to (\ref{U-spin-rel2}). Nevertheless, we observe 
that the impact of a variation of $\tilde\xi$ within $[0.8,1.2]$ is quite 
moderate. Interestingly, we may well accommodate values of $\alpha$ around 
$90^\circ$ in the case of $\phi_d=47^\circ$, in contrast to the situation of
$\phi_d=133^\circ$, which would favour $\alpha$ to lie around $40^\circ$.
Theoretical arguments for $\alpha\sim90^\circ$ were given in \cite{FX}.

Let us now compare our results with the analysis performed by 
the Belle collaboration in \cite{Belle-Bpipi}, yielding the range
\begin{equation}\label{Belle-range}
78^\circ\leq\phi_2\equiv\alpha\leq 152^\circ.
\end{equation}
The starting point of this study, which follows closely \cite{GR-Bpipi}, 
is also the parametrization (\ref{Bpipi-ampl}) of the $B^0_d\to\pi^+\pi^-$ 
decay amplitude. However, there are important differences in comparison
with our analysis:
\begin{enumerate}
\item[{\bf i)}]The range in (\ref{Belle-range}) corresponds to the 
95.5\% C.L. interval of the Belle asymmetries, 
constrained to lie within the physical region, whereas the results in 
Table~\ref{expval} were obtained for the averages of the BaBar and Belle 
measurements in (\ref{Bpipi-CP-averages}) and (\ref{Bpipi-CP-averages2})
(with ``ordinary'' errors), which fall well into the 95.5\% C.L. Belle 
region.
\item[{\bf ii)}] In the Belle analysis, $d$ was varied within $[0.15,0.45]$
to explore the penguin effects, whereas we use the observable $H$ to control 
the penguin contributions.
\item[{\bf iii)}] A crucial difference arises, since the SM relation 
$\phi_d=2\beta$ was used in \cite{Belle-Bpipi}, in combination with 
the unitarity relation $\beta+\gamma=\pi-\alpha$, to eliminate $\gamma$
(see (\ref{Bpipi-CP0})). Following these lines, the attention 
is implicitly restricted to the standard solution $\phi_d=47^\circ$
with $\beta=23.5^\circ$, so that the case of $\phi_d=133^\circ$ {\it cannot}
be explored.
\end{enumerate}
Despite the different input values for the CP asymmetries and 
the different treatment of the penguin contributions, 
(\ref{Belle-range}) is in good agreement with the ranges for $\alpha$ 
collected in Table~\ref{expval} for $\phi_d=47^\circ$.\footnote{If we
assume $\beta=23.5^\circ$ and insert the $\phi_d=47^\circ$ range for
$\gamma$ in (\ref{gam-res}) into $\alpha=180^\circ-\beta-\gamma$, we obtain 
$78^\circ\lsim\alpha\lsim122^\circ$. As (\ref{Belle-range}), this range 
does not depend on $R_b$, but relies on the SM relation $\phi_d=2\beta$.}
On the other hand, all the interesting features related to the 
$\phi_d=133^\circ$ solution were not addressed in the analysis in 
\cite{Belle-Bpipi}.

\boldmath
\subsection{Further Implications from $B\to\pi\pi, \pi K$}
\unboldmath
Interestingly, flavour-symmetry strategies to determine $\gamma$ with the
help of charged \cite{BpiK-NR} and neutral \cite{BpiK-BF} $B\to\pi K$ modes
show some preference for values of $\gamma$ larger than $90^\circ$
(for recent overviews, see \cite{neubert,BpiK-overviews}). Moreover, as
pointed out in \cite{RF-Bpipi,FlMa2}, the theoretical predictions
for the penguin parameter $de^{i\theta}$ obtained within the
``QCD factorization'' \cite{BBNS3} and ``PQCD'' \cite{PQCD-appl}
approaches can be brought to better agreement with the measured value of
$H$ in the case of $\gamma>90^\circ$. It should also be noted that the
global fits to all available $B\to\pi K$, $\pi\pi$ data show a similar
picture (see, for instance, \cite{BBNS3,HSW}). Since the $B\to\pi K$
analyses are rather involved, we shall not discuss them here in
further detail; let us just emphasize that their preferred values for
$\gamma$ could be conveniently accommodated in Fig.~\ref{fig:second}, i.e.\
for $\phi_d=133^\circ$, but not in the case of the conventional
$\phi_d=47^\circ$ solution shown in Fig.~\ref{fig:first}. Because of the
large experimental uncertainties, it is of course too early to draw definite
conclusions, but the experimental situation is expected to improve
continuously in the future. In this context, it is also helpful to construct
a set of robust sum rules \cite{sr}, which are satisfied by the $B\to\pi K$
observables.

\boldmath
\section{Implications for $K\to\pi\nu\overline{\nu}$ and
$B_{d,s}\to\mu^+\mu^-$}\label{sec:rare}
\unboldmath
In this section, we shall explore the implications of the allowed
regions in the $\overline{\rho}$--$\overline{\eta}$ plane shown
in Figs.~\ref{fig:first} and \ref{fig:second} for the branching
ratios of the very rare decay processes $K\to\pi\nu\overline{\nu}$
and $B_{d,s}\to\mu^+\mu^-$. Needless to note, the predictions
corresponding to the $\phi_d=133^\circ$ scenario are of particular
interest. As in the previous section, we assume that NP
does not affect the amplitudes of these modes, i.e.\ that it manifests
itself only through the different allowed ranges for $\overline{\rho}$ and
$\overline{\eta}$ shown in Figs.~\ref{fig:first} and \ref{fig:second}.
\boldmath
\subsection{$K\to\pi\nu\overline{\nu}$}\label{subsec:Kpinunu}
\unboldmath
As is well known, the rare kaon decays $K^+\to\pi^+\nu\overline{\nu}$
and $K_{\rm L}\to\pi^0\nu\overline{\nu}$ offer valuable tools to
explore flavour physics. These modes originate from $Z$ penguins and
box diagrams, and are theoretically very clean. Let us first focus
on $K^+\to\pi^+\nu\overline{\nu}$, which has already been observed by
the E787 experiment at BNL through two clean events, yelding the
following branching ratio \cite{E787}:
\begin{equation}\label{Kpnn-exp}
\mbox{BR}(K^+\to\pi^+\nu\overline{\nu})=\left(1.57^{+1.75}_{-0.82}\right)
\times10^{-10}.
\end{equation}
Within the SM and the NP scenarios specified
above, the ``reduced'' branching ratio \cite{BBSIN}
\begin{equation}\label{b1b2}
B_1=\frac{\mbox{BR}(K^+\to\pi^+\nu\overline{\nu})}{4.42\times 10^{-11}}
\end{equation}
takes the following form:
\begin{equation}\label{bkpn}
B_1=\left[\frac{\mbox{Im}\lambda_t}{\lambda^5}X(x_t)\right]^2+
\left[\frac{\mbox{Re}\lambda_t}{\lambda^5}X(x_t)+
\frac{\mbox{Re}\lambda_c}{\lambda} P_c(\nu\overline{\nu})\right]^2,
\end{equation}
where $X(x_t)=1.51 \pm 0.05$ and $P_c(\nu\overline{\nu})=0.40\pm0.06$
are appropriate coefficient functions, which encode top- and
charm-quark loop contributions, respectively \cite{BB98}. Taking into
account $\lambda_t\equiv V^\ast_{ts}V_{td}$ and
$\lambda_c\equiv V^\ast_{cs}V_{cd}$, we obtain
\begin{equation}
\mbox{Im}\lambda_t =  \eta A^2 \lambda^5,\quad
\mbox{Re}\lambda_t =-\left(1-\frac{\lambda^2}{2}\right)
A^2\lambda^5(1-\overline{\rho})
\end{equation}
and
\begin{equation}
\mbox{Re} \lambda_c =-\lambda
\left(1-\frac{\lambda^2}{2}\right),
\end{equation}
respectively, where $A\equiv |V_{cb}|/\lambda^2$.

The following ranges for the $K^+\to\pi^+\nu\overline{\nu}$ branching ratio 
are obtained by scanning $\overline{\rho}$ and $\overline{\eta}$ within the 
black regions shown in Figs.~\ref{fig:first} and \ref{fig:second}, and 
varying simultaneously $X(x_t)$, $P_c$ and the relevant parameters in 
Table~\ref{tab:inputs} within their allowed ranges. 
In the case when $\phi_d=47^\circ$, we obtain
\begin{equation}\label{Kpnn1}
0.33
\times
10^{-10}
\leq \mbox{BR}(K^+\to\pi^+\nu\overline{\nu})
\leq
1.19 \times 10^{-10},
\end{equation}
whereas the $\phi_d=133^\circ$ scenario favours a larger
branching ratio,
\begin{equation}\label{Kpnn2}
0.65
\times
10^{-10}
\leq \mbox{BR}(K^+\to\pi^+\nu\overline{\nu})
\leq
1.97 \times 10^{-10},
\end{equation}
which is due to the sign-change of $\overline{\rho}$ in (\ref{sym2}).
Despite the large present uncertainties, it is interesting to note that
(\ref{Kpnn2}) is in better agreement with the experimental result
(\ref{Kpnn-exp}), thereby favouring the $\phi_d=133^\circ$ solution.
On the other hand, (\ref{Kpnn1}) agrees well with the SM
range
\begin{equation}\label{Kpnn-SM}
\mbox{BR}(K^+\to\pi^+\nu\overline{\nu})=(0.72\pm0.21)\times 10^{-10}
\end{equation}
given in \cite{rising}, which is not surprising. Note, however,
that (\ref{Kpnn1}) relies only on the data on $\phi_d$, CP violation in
$B_d\to \pi^+\pi^-$, the observable $H$, and the measurement of the side
$R_b$ of the unitarity triangle. On the other hand, the usual CKM fits
were used in (\ref{Kpnn-SM}).

The observation that the central value of the  BNL-E787 result
does not necessarily imply a non-standard contribution to the 
$K^+\to\pi^+\nu\overline{\nu}$ amplitude, but that it could well be 
accommodated by  NP effects in
$B^0_d$--$\overline{B^0_d}$ mixing only,
has already been made in \cite{rising}.
Indeed, the SM interpretation of the
measured BR$(K^+\to\pi^+\nu\overline{\nu})$
defines a region in the $\overline{\rho}$--$\overline{\eta}$
plane which is perfectly consistent with the $R_b$ circle,
but favours $\gamma>90^\circ$. Adding to this analysis the 
$B_d \to \pi^+\pi^-$ information, we can now conclude that the 
measured $\mbox{BR}(K^+\to\pi^+\nu\overline{\nu})$ favours a 
$B^0_d$--$\overline{B^0_d}$ mixing phase of $\phi_d=133^\circ$.

Let us now have a brief look at the decay
$K_{\rm L}\to\pi^0\nu\overline{\nu}$, which is characterized by
\begin{equation}
B_2=\frac{\mbox{BR}(K_{\rm L}\to\pi^0\nu\overline{\nu})}{1.93\times 10^{-10}},
\end{equation}
with
\begin{equation}\label{bklpn}
B_2=\left[\frac{\mbox{Im}\lambda_t}{\lambda^5}X(x_t)\right]^2
=\eta^2A^4X(x_t)^2.
\end{equation}
In contrast to $B_1$, this reduced branching ratio does not depend on
$\overline{\rho}$. Consequently, since the allowed ranges for
$\overline{\eta}$ are equal in Figs.~\ref{fig:first} and \ref{fig:second}
because of (\ref{sym2}), $K_{\rm L}\to\pi^0\nu\overline{\nu}$ does not
allow us to distinguish between these two cases, i.e.\ we obtain the
same range for $\phi_d=47^\circ$ and $133^\circ$:
\begin{equation}\label{KLnn}
0.4 \times 10^{-11} \leq \mbox{BR}(K_{\rm
L}\to\pi^0\nu\overline{\nu}) \leq 6.2 \times 10^{-11},
\end{equation}
which overlaps well with the SM expectation \cite{Kettell}
\begin{equation}\label{KLnn-SM}
\mbox{BR}(K^+\to\pi^+\nu\overline{\nu})=(2.8\pm1.0)\times 10^{-11}.
\end{equation}

\boldmath
\subsection{$B_{d,s} \to \mu^+\mu^-$}
\unboldmath
Let us now, finally, turn to two rare $B$ decays, $B_d\to \mu^+\mu^-$ and
$B_s\to \mu^+\mu^-$. Within the SM and the kind of NP
specified above, they are mediated by box diagrams and $Z$ penguins.
We may write the branching ratio for $B_d\to \mu^+\mu^-$ as follows
\cite{Andr-Erice}:
\begin{eqnarray}
\mbox{BR}(B_d \to \mu^+ \mu^-) = 1.1 \times 10^{-10}
\left[ \frac{f_{B_d}}{0.20 \, \mbox{GeV}} \right]^2 \left[
\frac{|V_{td}|}{0.008}           \right]^2
\left[ \frac{\tau_{B_d}}{1.5 \, \mbox{ps}} \right] \left[
\frac{{\overline m}_t(m_t) }{167 \, \mbox{GeV} } \right]^{3.12},
\end{eqnarray}
where $|V_{td}|^2$ is given by
\begin{eqnarray}
|V_{td}|^2=A^2 \lambda^6R_t^2=A^2 \lambda^6
\left[(1-\overline{\rho})^2+\overline{\eta}^2\right].
\end{eqnarray}
We may now calculate, in analogy to our discussion of the 
$K\to\pi\nu\overline{\nu}$ modes given in Subsection~\ref{subsec:Kpinunu},
the range for this branching ratio corresponding to the black regions 
shown in Figs.~\ref{fig:first} and \ref{fig:second}. In the case of 
$\phi_d=47^\circ$, we obtain
\begin{equation}\label{Bdmumu1}
0.4 \times 10^{-10} \leq \mbox{BR}(B_d \to \mu^+ \mu^-) \leq 2.4
\times 10^{-10},
\end{equation}
whereas the $\phi_d=133^\circ$ scenario favours the following, larger
branching ratio:
\begin{equation}\label{Bdmumu2}
1.0 \times 10^{-10} \leq \mbox{BR}(B_d \to \mu^+ \mu^-) \leq 4.6
\times 10^{-10}.
\end{equation}
As in the case of $K^+\to\pi^+\nu\overline{\nu}$, the enhancement is due
to the sign-change of $\overline{\rho}$ in (\ref{sym2}).

On the other hand, we have
\begin{eqnarray}
\mbox{BR}( B_s \to \mu^+ \mu^-) = 4.1 \times 10^{-9} \left[
\frac{f_{B_s}}{0.24 \, \mbox{GeV}} \right]^2 \left[
\frac{|V_{ts}|}{0.040}           \right]^2 \left[
\frac{\tau_{B_s}}{1.5 \, \mbox{ps}} \right] \left[ \frac{{\overline
m}_t(m_t) }{167 \, \mbox{GeV} } \right]^{3.12},
\end{eqnarray}
where $V_{ts}$ is given -- up to tiny corrections entering at the
$\lambda^4$ level -- as follows:
\begin{equation}\label{Vts}
V_{ts}=-A\lambda^2=-V_{cb}.
\end{equation}
Consequently, since $\mbox{BR}( B_s \to \mu^+ \mu^-)$ does not depend
on $\overline{\rho}$ and $\overline{\eta}$, if we neglect these
corrections, it has no sensitivity on the allowed ranges in the
$\overline{\rho}$--$\overline{\eta}$ plane shown in Figs.~\ref{fig:first}
and \ref{fig:second}, i.e.\ we obtain the same prediction for
$\phi_d=47^\circ$ and $\phi_d=133^\circ$:
\begin{equation}\label{Bsmumu}
3 \times 10^{-9} \leq \mbox{BR}(B_s \to \mu^+ \mu^-) \leq 6
\times 10^{-9}.
\end{equation}
At first sight, $\mbox{BR}( B_s \to \mu^+ \mu^-)$ does therefore not appear
to be of great interest for our analysis. However, this is actually not
the case, since the ratio
\begin{equation}\label{mumu-rat}
\frac{\mbox{BR}(B_d \to \mu^+\mu^-)}{\mbox{BR}(B_s \to
\mu^+\mu^-)}=\left[\frac{\tau_{Bd}}{\tau_{B_s}}\right]\left[
\frac{M_{B_d}}{M_{B_s}}\right]\left[\frac{f_{B_d}}{f_{B_s}}\right]^2
\left|\frac{V_{td}}{V_{ts}}\right|^2
\end{equation}
is affected to a much smaller extent by the hadronic uncertainties
associated with $f_{B_d}$ and $f_{B_s}$ than the individual branching
ratios. These decay constants now enter in a ratio, which equals 1
in the $SU(3)$ limit, i.e.\ we have only to deal with the $SU(3)$-breaking
corrections to this quantity \cite{LL-02}:
\begin{equation}
\frac{f_{B_s}}{f_{B_d}}=1.18(4)^{+12}_{-0},
\end{equation}
whereas the ranges for the individual decay constants can be found in
Table~\ref{tab:inputs}. Using (\ref{Vts}) and taking (\ref{Rt-def})
into account, we obtain from (\ref{mumu-rat}) the following expression:
\begin{equation}
R\equiv \left[\frac{\tau_{Bs}}{\tau_{B_d}}\right]\left[
\frac{M_{B_s}}{M_{B_d}}\right]\left[\frac{f_{B_s}}{f_{B_d}}\right]^2
\left[\frac{\mbox{BR}(B_d \to \mu^+\mu^-)}{\mbox{BR}(B_s \to
\mu^+\mu^-)}\right]=\lambda^2 R_t^2.
\end{equation}
Since the side $R_t$ of the unitarity triangle takes quite different values
for the situations shown in Figs.~\ref{fig:first} and \ref{fig:second}
because of (\ref{sym2}), we may nicely probe them through $R$. In the
case of $\phi_d=47^\circ$, we find
\begin{equation}
2.2 \times 10^{-2} \leq R
\leq 5.4 \times 10^{-2},
\end{equation}
while our $\phi_d=133^\circ$ scenario corresponds to
\begin{equation}
5.7 \times 10^{-2} \leq
R \leq
10.1 \times 10^{-2}.
\end{equation}
Unfortunately, the sensitivity of the existing experiments is still far
from the level necessary to probe the ratio $R$. The present
experimental upper bounds read as follows:
\begin{equation}
\mbox{BR}( B_s \to \mu^+ \mu^-) < 2.6\times10^{-6} \quad\mbox{(95\% C.L.
\cite{CDF-rare})}
\end{equation}
\begin{equation}
\mbox{BR}( B_d \to \mu^+ \mu^-) < 2.0\times10^{-7} \quad\mbox{(90\% C.L.
\cite{BABAR-rare}).}
\end{equation}
However, owing to clean experimental signatures, these
processes are among the benchmark modes of future
$B$-physics experiments at hadron colliders.

\section{Conclusions and Outlook}\label{sec:concl}
The main points of this paper can be summarized as follows:
\begin{itemize}
\item We have emphasized that the CP-violating $B^0_d$--$\overline{B^0_d}$
mixing phase $\phi_d\sim 47^\circ\lor 133^\circ$, which has been
determined from ${\cal A}_{\rm CP}^{\rm mix}(B_d\to J/\psi K_{\rm S})$,
cannot be represented straightforwardly in the
$\overline{\rho}$--$\overline{\eta}$ plane in the presence of
NP contributions to $B^0_d$--$\overline{B^0_d}$ mixing. This
feature affects in particular the ``unconventional'' solution of
$\phi_d=133^\circ$, which definitely requires such kind of physics
beyond the SM. We have pointed out that it is therefore very misleading
to represent this solution simply as a second branch
in the $\overline{\rho}$--$\overline{\eta}$ plane, as is usually done in
the literature.

\item In the solution of this problem, $R_b$ and $\gamma$ play a key
r\^ole. From a theoretical point of view, certain pure tree decays,
for example $B_d\to D K_{\rm S}$ or $B_s\to D\phi$ modes, would offer
an ideal tool to determine $\gamma$, providing theoretically clean results
that are, in addition, very robust under the influence of NP.
Unfortunately, these strategies cannot yet be implemented. However, an
interesting alternative is offered by the $B$-factory benchmark mode
$B_d\to\pi^+\pi^-$, where first experimental results on CP-violating effects
are already available from BaBar and Belle.

\item We have considered a specific scenario for physics beyond the
SM, where we have large NP contributions to
$B^0_d$--$\overline{B^0_d}$ mixing, but not to the $\Delta B=1$ and
$\Delta S=1$ decay processes. Within this framework, we may then also
straightforwardly accommodate the $\phi_d=133^\circ$ solution. We have
given general conditions for such a kind of NP, and have argued
that it is well motivated in supersymmetry and several specific frameworks.

\item As widely discussed in the literature, the CP-violating 
$B_d\to\pi^+\pi^-$ observables provide valuable information about 
the ``true'' apex of the unitarity triangle for such a scenario 
of NP. However, the true challenge in this game is the control of 
penguin effects. To this purpose, we employ the CP-averaged 
$B_d\to\pi^\pm K^\mp$ branching ratio, 
and make use of $U$-spin and plausible dynamical assumptions. Following 
these lines, we may determine $\gamma$ for each of the two solutions for 
$\phi_d$. Complementing this information with the experimental range for 
$R_b$, we may fix the apex of the unitarity triangle, and thus determine
also $\alpha$ and $\beta$. In the case of $\phi_d\sim47^\circ$, we 
arrive at the first quadrant of the $\overline{\rho}$--$\overline{\eta}$ 
plane, and obtain an allowed region, which is in perfect agreement with 
the constraints arising from the SM interpretation of $\phi_d$, $\Delta M_d$ 
and $\Delta M_s/\Delta M_d$, as well as of $\varepsilon_K$. On the other hand, 
for $\phi_d\sim133^\circ$, we obtain an allowed region in the second 
quadrant, corresponding to $\gamma>90^\circ$. Interestingly, this range 
is still consistent with the $\varepsilon_K$ hyperbola, whereas we may 
now no longer use $\Delta M_d$ and $\Delta M_s/\Delta M_d$ to determine 
$R_t$, as these quantities would receive NP contributions.
               
\item We have also explored the implications of these two solutions for
very rare $K$ and $B$ decays, $K\to\pi\nu\overline{\nu}$ and
$B_{d,s}\to\mu^+\mu^-$, respectively, assuming that their decay amplitudes
take the same form as in the SM. In this case, NP
would manifest itself only indirectly, through the determination of
$\overline{\rho}$ and $\overline{\eta}$.
Consequently, for $\phi_d=47^\circ$ we obtain the same picture for these
modes as in the SM. However, in the case of $\phi_d\sim133^\circ$, the
branching ratio for $K^+\to\pi^+\nu\overline{\nu}$ that would be favoured
is about twice as large as in the SM, and would be in better agreement
with the present measurement. Similarly, if $\phi_d\sim133^\circ$, 
a branching ratio about twice as large as in the SM is expected for 
$B_d\to\mu^+\mu^-$. On the other hand, the impact of this
non-standard solution on both BR$(B_s\to\mu^+\mu^-)$ and
BR$(K_{\rm L}\to\pi^0\nu\overline{\nu})$ would be negligible.
\end{itemize}

\noindent
At the present stage, we may not yet draw any definite conclusion, because
of the large experimental uncertainties and the unsatisfactory situation
of the measurement of the CP-violating $B_d\to\pi^+\pi^-$ observables;
hopefully, the discrepancy between BaBar and Belle will be resolved soon.
Interestingly, alternative analyses of the data for the $B\to\pi K, \pi\pi$
branching ratios show some preference for $\gamma>90^\circ$, thereby
favouring the $\phi_d=133^\circ$ scenario. A similar comment applies to
the rare kaon decay $K^+\to\pi^+\nu\overline{\nu}$.
On the other hand, it should not be forgotten that the allowed region for 
$\phi_d=47^\circ$ is consistent with the stringent constraints from 
$B^0_{d,s}$--$\overline{B^0_{d,s}}$ mixing. As far as $\varepsilon_K$ is 
concerned, both the $\phi_d=47^\circ$ and the $\phi_d=133^\circ$ scenario 
are consistent with the SM interpretation of this quantity. In view of 
these observations, it would be very important to distinguish directly 
between the two solutions for $\phi_d$, which are implied by their 
extraction from $\sin\phi_d$, through a  measurement of the sign of 
$\cos\phi_d$.

\vspace*{0.6truecm}

\noindent
{\it Acknowledgements}

\vspace*{0.3truecm}

\noindent
J.M. acknowledges financial support by MCyT and FEDER FPA2002-00748.  
The work by G.I. and J.M. is partially supported by IHP-RTN, EC 
contract No.\ HPRN-CT-2002-00311 (EURIDICE).

%%%%%%%%%%%%%%%%%%%%%%%%%%%%%%%%%%%%%%%%%%%%%%%%%%%%%%%%%%%%%%%%%%%%%%%%%%%%


\begin{thebibliography}{99}
%%%%%%%%%%%%%%%%%%%%%%%%%%%%%%%%%%%%%%%%%%%%%%%%%%%%%%%%%%%%%%%%%%%%%%%%%%%%


\bibitem{KM}M. Kobayashi and T. Maskawa,
%``CP Violation In The Renormalizable Theory Of Weak Interaction,''
{\it Prog.\ Theor.\ Phys.}~{\bf 49} (1973) 652.
%%CITATION = PTPKA,49,652;%%

\bibitem{CP-B-obs}BaBar Collaboration (B. Aubert {\it et al.}),
%``Observation of CP violation in the B0 meson system,''
{\it Phys.\ Rev.\ Lett.}~{\bf 87} (2001) 091801;\\
%%CITATION = HEP-EX 0107013;%%
Belle Collaboration (K. Abe {\it et al.}),
%``Observation of large CP violation in the neutral B meson system,''
{\it Phys.\ Rev.\ Lett.}~{\bf 87} (2001) 091802.
%%CITATION = HEP-EX 0107061;%%

\bibitem{BLO}L. Wolfenstein, {\it Phys.\ Rev.\ Lett.}~{\bf 51} (1983) 1945;\\
%%CITATION = PRLTA,51,1945;%%
A.J. Buras, M.E. Lautenbacher and G. Ostermaier,
%``Waiting for the top quark mass, K+ $\to$ pi+ neutrino anti-neutrino,
%B(s)0 - anti-B(s)0 mixing and CP asymmetries in B decays,''
{\it Phys.\ Rev.}~{\bf D50} (1994) 3433.
%%CITATION = HEP-PH 9403384;%%

\bibitem{CKM-fits}A.J. Buras, TUM-HEP-435-01 [hep-ph/0109197];\\
%``Flavour physics and CP violation in the SM,''
%%CITATION = HEP-PH 0109197;%%
A. Ali and D. London,
%``What if the mass difference Delta(M(s)) is around 18 inverse
%picoseconds?,''
{\it Eur.\ Phys.\ J.}~{\bf C18} (2001) 665;\\
%%CITATION = HEP-PH 0012155;%%
D. Atwood and A. Soni,
%``CP asymmetry measurements in psi K0 and the CKM paradigm,''
{\it Phys.\ Lett.}~{\bf B508} (2001) 17;\\
%%CITATION = HEP-PH 0103197;%%
M.~Ciuchini {\it et al.},
%``2000 CKM-triangle analysis: A critical review with updated experimental
%inputs and theoretical parameters,''
{\it JHEP} {\bf 0107} (2001) 013;\\
%%CITATION = HEP-PH 0012308;%%
A. H\"ocker {\it et al.},
%``A new approach to a global fit of the CKM matrix,''
{\it Eur.\ Phys.\ J.}~{\bf C21} (2001) 225.
%%CITATION = HEP-PH 0104062;%%

\bibitem{nir} Y. Nir, WIS-35-02-DPP [hep-ph/0208080].
%``CP violation: The CKM matrix and new physics,''
%%CITATION = HEP-PH 0208080;%%

\bibitem{RF-PHYS-REP}R. Fleischer,
%``CP violation in the B system and relations to K $\to$ pi nu anti-nu
%decays,''
{\it Phys.\ Rep.}~{\bf 370} (2002) 537.
%%CITATION = HEP-PH 0207108;%%

\bibitem{FM-BpsiK}R. Fleischer and T. Mannel,
%``General analysis of new physics in B $\to$ J/psi K,''
{\it Phys.\ Lett.}~{\bf B506} (2001) 311.
%%CITATION = HEP-PH 0101276;%%

\bibitem{rising}G. D'Ambrosio and G. Isidori,
{\it Phys.\ Lett.}~{\bf B530} (2002) 108.
%%CITATION = HEP-PH 0112135;%%

\bibitem{Becirevic}D. Becirevic {\it et al.},
{\it Nucl.\ Phys.}~{\bf B634} (2002) 105.
%%CITATION = HEP-PH 0112303;%%

\bibitem{GNW}Y. Grossman, Y. Nir and M.P. Worah,
%``A model independent construction of the unitarity triangle,''
{\it Phys.\ Lett.}~{\bf B407} (1997) 307.
%%CITATION = HEP-PH 9704287;%%

\bibitem{Univ_tri}A.J. Buras {\it et al.},
{\it Phys.\ Lett.}~{\bf B500} (2001) 161.
%%CITATION = HEP-PH 0007085;%%

\bibitem{DGIS}G. D'Ambrosio, G.F. Giudice, G. Isidori and A. Strumia,
{\it Nucl.\ Phys.}~{\bf B645} (2002) 155.
%%CITATION = HEP-PH 0207036;%%

\bibitem{RF-Bpipi}R. Fleischer,
%``Constraining penguin contributions and the CKM angle gamma through
%B/d $\to$ pi+ pi-,''
{\it Eur.\ Phys.\ J.}~{\bf C16} (2000) 87.
%%CITATION = HEP-PH 0001253;%%

\bibitem{FlMa2}R. Fleischer and J. Matias,
%``Exploring CP violation through correlations in B $\to$ pi K,
%B/d $\to$ pi+ pi-, B/s $\to$ K+ K- observable space,''
{\it Phys.\ Rev.}~{\bf D66} (2002) 054009.
%%CITATION = HEP-PH 0204101;%%

\bibitem{BBNS3}M. Beneke {\it et al.},
%``QCD factorization in B $\to$ pi K, pi pi decays and extraction of
%Wolfenstein parameters,''
{\it Nucl.\ Phys.}~{\bf B606} (2001) 245.
%%CITATION = HEP-PH 0104110;%%

\bibitem{neubert}M. Neubert, CLNS-02-1794 [hep-ph/0207327].
%``Two-body modes of B mesons and the CP-b triangle,''
%%CITATION = HEP-PH 0207327;%%

\bibitem{alt-strat}J.P. Silva and L. Wolfenstein,
%``Determining the penguin effect on CP violation in B0 $\to$ pi+ pi-,''
{\it Phys.\ Rev.}~{\bf D49} (1994) 1151;\\
%%CITATION = HEP-PH 9309283;%%
A.L. Kagan and M. Neubert,
%``Implications of a low sin(2beta): A strategy for exploring new flavor
%physics,''
{\it Phys.\ Lett.}~{\bf B492} (2000) 115;\\
%%CITATION = HEP-PH 0007360;%%
G. Eyal, Y. Nir and G. Perez,
%``Implications of a small CP asymmetry in B $\to$ psi K(S),''
{\it JHEP} {\bf 0008} (2000) 028;\\
%%CITATION = HEP-PH 0008009;%%
F. Botella, G. Branco, M. Nebot and M. Rebelo,
%``Unitarity triangles and the search for new physics,''
{\it Nucl.\ Phys.}~{\bf B651} (2003) 174;\\
%%CITATION = HEP-PH 0206133;%%
A.J. Buras, F. Parodi and A. Stocchi,
%``The CKM matrix and the unitarity triangle: Another look,''
{\it JHEP} {\bf 0301} (2003) 029.
%%CITATION = HEP-PH 0207101;%%

\bibitem{ambig}Ya.I. Azimov, V.L. Rappoport and V.V. Sarantsev,
{\it Z. Phys.}~{\bf A356} (1997) 437;\\
%%CITATION = HEP-PH 9608478;%%
Y. Grossman and H.R. Quinn, {\it Phys.\ Rev.}~{\bf D56} (1997) 7259;\\
%%CITATION = HEP-PH 9705356;%%
J. Charles {\it et al.}, {\it Phys.\ Lett.}~{\bf B425} (1998) 375;\\
%%CITATION = HEP-PH 9801363;%%
B. Kayser and D. London, {\it Phys.\ Rev.}~{\bf D61} (2000) 116012;\\
%%CITATION = HEP-PH 9909560;%%
H.R. Quinn {\it et al.}, {\it Phys.\ Rev.\ Lett.}~{\bf 85} (2000) 5284.
%%CITATION = HEP-PH 0008021;%%

\bibitem{DDF2DFN}A.S. Dighe, I. Dunietz and R. Fleischer,
%``Resolving a discrete ambiguity in the CKM angle beta through
%B(u,d) $\to$ J/psi K* and B/s $\to$ J/psi Phi decays,''
{\it Phys.\ Lett.}~{\bf B433} (1998) 147;\\
%%CITATION = HEP-PH 9804254;%%
I. Dunietz, R. Fleischer and U. Nierste,
%``In pursuit of new physics with B/s decays,''
{\it Phys.\ Rev.}~{\bf D63} (2001) 114015.
%%CITATION = HEP-PH 0012219;%%

\bibitem{itoh}R. Itoh, KEK-PREPRINT-2002-106 [hep-ex/0210025].
%``Measurements of time dependent CP asymmetry in B $\to$ V V decays with
%BELLE,''
%%CITATION = HEP-EX 0210025;%%

\bibitem{tree-gam}M. Gronau and D. Wyler,
%``On determining a weak phase from CP asymmetries in charged B decays,''
{\it Phys.\ Lett.}~{\bf B265} (1991) 172;\\
%%CITATION = PHLTA,B265,172;%%
D. Atwood, I. Dunietz and A. Soni,
%``Enhanced CP violation with B $\to$ K D0 (anti-D0) modes and extraction
%of the CKM angle gamma,''
{\it Phys.\ Rev.\ Lett.}~{\bf 78} (1997) 3257;\\
%%CITATION = HEP-PH 9612433;%%
R. Fleischer and D. Wyler,
%``Exploring CP violation with B/c decays,''
{\it Phys.\ Rev.}~{\bf D62} (2000) 057503.\\
%%CITATION = HEP-PH 0004010;%%
Y. Grossman, Z. Ligeti and A. Soffer, LBNL-51630 [hep-ph/0210433];\\
%``Measuring gamma in B+- $\to$ K+- (K K*)(D) decays,''
%%CITATION = HEP-PH 0210433;%%
M. Gronau, CERN-TH/2002-331 [hep-ph/0211282].
%``Improving bounds on gamma in B+- $\to$ D K+- and B+-,0 $\to$ D X/s+-,0,''
%%CITATION = HEP-PH 0211282;%%

\bibitem{D-mix-incl}J.P. Silva and A. Soffer,
%``Impact of D0 anti-D0 mixing on the experimental determination of gamma,''
{\it Phys.\ Rev.}~{\bf D61} (2000) 112001;\\
%%CITATION = HEP-PH 9912242;%%
D. Atwood, I. Dunietz and A. Soni,
%``Improved methods for observing CP violation in B+- $\to$ K D and
%measuring the CKM phase gamma,''
{\it Phys.\ Rev.}~{\bf D63} (2001) 036005.
%%CITATION = HEP-PH 0008090;%%

\bibitem{RF-BDfr1}R. Fleischer, CERN-TH/2003-010, to appear in
{\it Phys.\ Lett.}~{\bf B} [hep-ph/0301255].
%%CITATION = HEP-PH 0301255;%%

\bibitem{RF-BDfr2}R. Fleischer, CERN-TH/2003-011, to appear in
{\it Nucl.\ Phys.}~{\bf B}  [hep-ph/0301256].
%%CITATION = HEP-PH 0301256;%%

\bibitem{RF-BsKK}R. Fleischer,
%``New strategies to extract beta and gamma from B/d $\to$ pi+ pi- and
%B/s $\to$ K+ K-,''
{\it Phys.\ Lett.}~{\bf B459} (1999) 306.
%%CITATION = HEP-PH 9903456;%%

\bibitem{BF2001}A.J. Buras and R. Fleischer,
{\it Phys.\ Rev.}~{\bf D64} (2001) 115010.
%%CITATION = HEP-PH 0104238;%%

\bibitem{BpiK-NP}M. Ciuchini, E. Franco, A. Masiero and L. Silvestrini,
%``b $\to$ s transitions: A new frontier for indirect SUSY searches,''
hep-ph/0212397; \\
%%CITATION = HEP-PH 0212397;%%
R. Harnik, D.T. Larson, H. Murayama and A. Pierce,
%``Atmospheric neutrinos can make beauty strange,''
hep-ph/0212180.
%%CITATION = HEP-PH 0212180;%%

\bibitem{FM-BphiK}R. Fleischer and T. Mannel,
%``Exploring new physics in the B $\to$ Phi K system,''
{\it Phys.\ Lett.}~{\bf B511} (2001) 240.
%%CITATION = HEP-PH 0103121;%%

\bibitem{GGMS96}J.S. Hagelin, S. Kelley and T. Tanaka,
{\it Nucl.\ Phys.}~{\bf B415} (1994) 293;\\
%%CITATION = NUPHA,B415,293;%%
F. Gabbiani {\it et al.},
{\it Nucl.\ Phys.}~{\bf B477} (1996) 321.
%%CITATION = HEP-PH 9604387;%%

\bibitem{BBL}G. Buchalla, A.J. Buras and M.E. Lautenbacher,
%``Weak Decays Beyond Leading Logarithms,''
{\it Rev.\ Mod.\ Phys.}~{\bf 68} (1996) 1125.
%%CITATION = HEP-PH 9512380;%%

\bibitem{RGEb}D. Choudhury {\it et al.}, 
{\it Phys.\ Lett.}~{\bf B342} (1995) 180.
%%CITATION = HEP-PH 9408275;%%

\bibitem{Lunghi}E. Lunghi, A. Masiero, I. Scimemi and L. Silvestrini,
{\it Nucl.\ Phys.}~{\bf B568} (2000) 120.
%%CITATION = HEP-PH 9906286;%%

\bibitem{BRS}A.J. Buras, A. Romanino and L. Silvestrini,
%``K $\to$ pi nu anti-nu: A model independent analysis and supersymmetry,''
{\it Nucl.\ Phys.}~{\bf B520} (1998) 3.
%%CITATION = HEP-PH 9712398;%%

\bibitem{BS}R. Barbieri and A. Strumia,
{\it Nucl.\ Phys.}~{\bf B508} (1997) 3.
%%CITATION = HEP-PH 9704402;%%

\bibitem{CIMM} G. Colangelo and G. Isidori, {\it JHEP}~{\bf 09} (1998) 009;\\
%%CITATION = HEP-PH 9808487;%%
A. Masiero and H. Murayama, {\it Phys.\ Rev.\ Lett.}~{\bf 83} (1999)
907.
%%CITATION = HEP-PH 9903363;%%

\bibitem{BCIRS}A.J. Buras {\it et al.},
%``Connections between epsilon'/epsilon and rare kaon decays in  
%supersymmetry,''
{\it Nucl.\ Phys.}~{\bf B566} (2000) 3.
%%CITATION = HEP-PH 9908371;%%

\bibitem{BK}K.S. Babu and C.F. Kolda,
%``Higgs-mediated B0 $\to$ mu+ mu- in minimal supersymmetry,''
{\it Phys.\ Rev.\ Lett.}~{\bf 84} (2000) 228; \\
%%CITATION = HEP-PH 9909476;%%
A.J. Buras {\it et al.},
% ``Delta(M(s))/Delta(M(d)), sin 2beta and the angle gamma in the
% presence  of new Delta(F) = 2 operators,''
{\it Nucl.\ Phys.}~{\bf B619} (2001) 434; \\
%%CITATION = HEP-PH 0107048;%%
G. Isidori and A. Retico,
%``Scalar flavour-changing neutral currents in the large-tan(beta) limit,''
{\it JHEP} {\bf 0111} (2001) 001.
%%CITATION = HEP-PH 0110121;%%

\bibitem{BaBar-Bpipi}B. Aubert {\it et al.}\ (BaBar Collaboration),
%``Measurements of branching fractions and CP-violating asymmetries in B0
%$\to$ pi+ pi-, K+ pi-, K+ K- decays,''
{\it Phys.\ Rev.\ Lett.}~{\bf 89} (2002) 281802.
%%CITATION = HEP-EX 0207055;%%

\bibitem{Belle-Bpipi}K. Abe {\it et al.}\ (Belle Collaboration),
Belle preprint 2003-1 [hep-ex/0301032].
%%CITATION = HEP-EX 0301032;%%

\bibitem{PDG}Particle Data Group (K. Hagiwara {\it et al.}),
%``Review Of Particle Physics,''
{\it Phys.\ Rev.}~{\bf D66} (2002) 010001.
%%CITATION = PHRVA,D66,010001;%%

\bibitem{GR-Bpipi}M. Gronau and J.L. Rosner,
%``Strong and weak phases from time-dependent measurements of B $\to$ pi pi,''
{\it Phys.\ Rev.}~{\bf D65} (2002) 093012.
%%CITATION = HEP-PH 0202170;%%

\bibitem{PQCD-appl}A.I. Sanda and K. Ukai,
%``How well can we predict CP asymmetry in B $\to$ pi pi, K pi decays?,''
{\it Prog.\ Theor.\ Phys.}~{\bf 107} (2002) 421;\\
%%CITATION = HEP-PH 0109001;%%
Y.-Y. Keum, DPNU-02-30 [hep-ph/0209208]
%%CITATION = HEP-PH 0209208;%%

\bibitem{Bpipi-strategies}M. Gronau and D. London,
%``Isospin Analysis Of CP Asymmetries In B Decays,''
{\it Phys.\ Rev.\ Lett.}~{\bf 65} (1990) 3381;\\
%%CITATION = PRLTA,65,3381;%%
R. Fleischer and T. Mannel,
%``Penguin zoology in B $\to$ pi pi and the extraction of the CKM angle
%alpha,''
{\it Phys.\ Lett.}~{\bf B397} (1997) 269;\\
%%CITATION = HEP-PH 9610357;%%
Y. Grossman and H.R. Quinn,
%``Bounding the effect of penguin diagrams in a(CP)(B0 $\to$ pi+ pi-),''
{\it Phys.\ Rev.}~{\bf D58} (1998) 017504;\\
%%CITATION = HEP-PH 9712306;%%
J. Charles,
%``Taming the penguin in the B/d0(t) $\to$ pi+ pi- CP-asymmetry:
%Observables and minimal theoretical input,''
{\it Phys.\ Rev.}~{\bf 59} (1999) 054007; \\
%%CITATION = HEP-PH 9806468;%%
D. London, N. Sinha and R. Sinha,
%``Searching for new physics via CP violation in B $\to$ pi pi,''
{\it Phys.\ Rev.}~{\bf D63} (2001) 054015;\\
%%CITATION = HEP-PH 0010174;%%
M. Gronau, D. London, N. Sinha and R. Sinha,
%``Improving bounds on penguin pollution in B $\to$ pi pi,''
{\it Phys.\ Lett.}~{\bf B514} (2001) 315;\\
%%CITATION = HEP-PH 0105308;%%
M. Gronau and J.L. Rosner,
%``Strong and weak phases from time-dependent measurements of B $\to$ pi pi,''
{\it Phys.\ Rev.}~{\bf D66} (2002) 053003.

\bibitem{TEV-Report}K. Anikeev {\it et al.}, FERMILAB-Pub-01/197
%``B physics at the Tevatron: Run II and beyond,''
[hep-ph/0201071].
%%CITATION = HEP-PH 0201071;%%

\bibitem{LHC-Report}P. Ball {\it et al.}, CERN-TH/2000-101
[hep-ph/0003238], in CERN Report on Standard Model physics (and more) at
the LHC (CERN, Geneva, 2000) p.\ 305.
%``B decays at the LHC,''
%%CITATION = HEP-PH 0003238;%%

\bibitem{CLEO-BpiK}CLEO Collaboration (A. Bornheim {\it et al.}),
CLEO 03-03 [hep-ex/0302026].
%%CITATION = HEP-EX 0302026;%%

\bibitem{BaBar-BpiK}BaBar Collaboration (B. Aubert {\it et al.}),
hep-ex/0207065; hep-ex/0206053.

\bibitem{Belle-BpiK}Belle Collaboration (B.C. Casey {\it et al.}),
%``Charmless hadronic two-body B meson decays,''
{\it Phys.\ Rev.}~{\bf D66} (2002) 092002.
%%CITATION = HEP-EX 0207090;%%

\bibitem{tomura}T. Tomura, talk at {\tt http://moriond.in2p3.fr/EW/2003/}

\bibitem{FlMa1}R. Fleischer and J. Matias,
%``Searching for new physics in non-leptonic B decays,''
{\it Phys.\ Rev.}~{\bf D61} (2000) 074004.
%%CITATION = HEP-PH 9906274;%%

\bibitem{GKN}Y. Grossman, B. Kayser and Y. Nir,
%``The role of the vacuum insertion approximation in calculating CP
%asymmetries in B decays,''
{\it Phys.\ Lett.}~{\bf B415} (1997) 90.
%%CITATION = HEP-PH 9708398;%%

\bibitem{FX}H. Fritzsch and Z.-Z. Xing,
%``The light quark sector, CP violation, and the unitarity triangle,''
{\it Nucl.\ Phys.}~{\bf B556} (1999) 49.
%%CITATION = HEP-PH 9904286;%%

\bibitem{BpiK-NR}M. Neubert and J.L. Rosner,
%``New bound on gamma from B+- $\to$ pi K decays,''
{\it Phys.\ Lett.}~{\bf B441} (1998) 403;
%%CITATION = HEP-PH 9808493;%%
%``Determination of the weak phase gamma from rate measurements in
%B+- $\to$ pi K, pi pi decays,''
{\it Phys.\ Rev.\ Lett.}~{\bf 81} (1998) 5076.
%%CITATION = HEP-PH 9809311;%%

\bibitem{BpiK-BF}A.J. Buras and R. Fleischer,
%``Constraints on the CKM angle gamma and strong phases from B $\to$ pi K
%decays,''
{\it Eur.\ Phys.\ J.}~{\bf C16} (2000) 97;
%%CITATION = HEP-PH 0003323;%%
%``A general analysis of gamma determinations from B $\to$ pi K decays,''
{\bf C11} (1999) 93.
%%CITATION = HEP-PH 9810260;%%

\bibitem{BpiK-overviews}J.L. Rosner, EFI-02-96 [hep-ph/0207197];\\
%%CITATION = HEP-PH 0207197;%%
R. Fleischer, CERN-TH/2002-293 [hep-ph/0210323].
%``B physics and CP violation,''
%%CITATION = HEP-PH 0210323;%%

\bibitem{HSW}W.S. Hou, J.G. Smith and F. W\"urthwein,
%``Determination of the phase of V(ub) from charmless hadronic B decay
%rates,''
NTU-HEP-99-25 [hep-ex/9910014].
%%CITATION = HEP-EX 9910014;%%

\bibitem{sr}M. Gronau and J.L. Rosner,
%``Combining CP asymmetries in B $\to$ K pi decays,''
{\it Phys.\ Rev.}~{\bf D59} (1999) 113002;\\
%%CITATION = HEP-PH 9809384;%%;
M. Neubert,
%``Model-independent analysis of B $\to$ pi K decays and bounds
% on the weak  phase gamma,''
{\it JHEP} {\bf 9902} (1999) 014;\\
%%CITATION = HEP-PH 9812396;%%;
J. Matias,
%``Model independent sum rules for B $\to$ pi K decays,''
{\it Phys.\ Lett.}~{\bf B520} (2001) 131 and hep-ph/0209331.
%%CITATION = HEP-PH 0105103;%%
%%CITATION = HEP-PH 0209331;%%

\bibitem{E787}E787 Collaboration (S. Adler {\it et al.}),
%``Further evidence for the decay K+ $\to$ pi+ nu anti-nu,''
{\it Phys.\ Rev.\ Lett.}~{\bf 88} (2002) 041803.
%%CITATION = HEP-EX 0111091;%%

\bibitem{BBSIN}G. Buchalla and A.J. Buras,
{\it Phys.\ Lett.}~{\bf B333} (1994) 221;
{\it Phys.\ Rev.}~{\bf D54} (1996) 6782.

\bibitem{BB98}G. Buchalla and A.J. Buras,
{\it Nucl.\ Phys.}~{\bf B548} (1999) 309; {\bf B412} (1994) 106.

\bibitem{Kettell}S.H. Kettell, L.G. Landsberg and H.H. Nguyen,
%``Estimate of B(K $\to$ pi nu anti-nu)$|$SM using the kaon unitary triangle,''
FERMILAB-FN-0727 [hep-ph/0212321].
%%CITATION = HEP-PH 0212321;%%

\bibitem{Andr-Erice}A.J. Buras,
%``Flavor dynamics: CP violation and rare decays,''
TUM-HEP-402-01 [hep-ph/0101336].
%%CITATION = HEP-PH 0101336;%%

\bibitem{LL-02}L. Lellouch,
%``Phenomenology from lattice QCD,''
CPT-2002-P-4443 [hep-ph/0211359].
%%CITATION = HEP-PH 0211359;%%

\bibitem{CDF-rare}CDF Collaboration (F. Abe {\it et al.}),
%``Search for the decays B/d0 $\to$ mu+ mu- and B/s0 $\to$ mu+ mu- in
%p anti-p collisions at s**(1/2) = 1.8-TeV,''
{\it Phys.\ Rev.}~{\bf D57} (1998) 3811.
%%CITATION = PHRVA,D57,3811;%%

\bibitem{BABAR-rare}BaBar Collaboration (B. Aubert {\it et al.}),
BABAR-CONF-02/028 [hep-ex/0207083].

\end{thebibliography}
\end{document}